\documentclass[lettersize,journal]{IEEEtran}

\usepackage{amsmath,amsfonts,amssymb}
\usepackage{array}
\usepackage{booktabs}
\usepackage{multirow}
\usepackage{makecell}
\usepackage{graphicx}
\usepackage{subcaption}
\usepackage{textcomp}
\usepackage{stfloats}
\usepackage{url}
\usepackage{cite}

\begin{document}

\title{TG-DIN: Theory-Guided Demand Inference Network for Generalizable QoS Prediction}

\author{Fuliang~Yang~and~Feng~Ye
\thanks{The authors are with the Department of Electrical and Computer Engineering, University of Wisconsin-Madison, Madison, WI 53706 USA (e-mail: fuliang.yang@wisc.edu; feng.ye@wisc.edu).}%
\thanks{This project is partially supported by the U.S. National Science Foundation under grants 2344341 and 2542859.}}

\markboth{Under Review}%
{Yang \MakeLowercase{\textit{et al.}}: TG-DIN: Theory-Guided Demand Inference Network}

\maketitle

\begin{abstract}
Network measurements reveal the service a flow receives, not the load it offers to the network. Predictors trained directly on observed traffic therefore become tied to the operating conditions in their training data and can degrade when bottleneck capacity or traffic behavior changes. We present TG-DIN, a theory-guided framework that infers the latent demand each flow offers to a shared bottleneck from observable quality-of-service (QoS) signals. At its core is a differentiable theory layer that encodes bottleneck scheduling and queuing and serves two roles. Run forward on sampled demand and network conditions, it generates randomized, mechanism-consistent training traces; placed after a set-based demand predictor, it maps inferred demand to throughput, delay, and loss. Training therefore relies only on reconstructing observable QoS and requires neither demand labels nor collected traffic. Because bottleneck capacity enters the theory layer as an input, a single model serves bottlenecks of different capacities. In synthetic experiments spanning 20 to 360~Mbps, TG-DIN is far more robust to capacity shifts than direct data-driven predictors, with a per-flow throughput RMSE of 13.65~Mbps at 360~Mbps against 71.45 to 82.74~Mbps, and it outperforms predictors fine-tuned with up to 5\% target-capacity data at every tested capacity. Although trained entirely on generated traces, TG-DIN transfers without fine-tuning to real testbed measurements, including dynamic capacity changes and a switch from video streaming to a TCP bulk download. A complete training run takes about 36 minutes on a single GPU.
\end{abstract}


\IEEEpeerreviewmaketitle

\section{Introduction}

Network users and operators observe quality-of-service (QoS) metrics such as throughput, delay, and packet loss. These metrics capture the outcomes of network operation but not the conditions that produced them. Low throughput, for example, may result from low demand, limited capacity, contention with other flows, or scheduling effects. We therefore treat each flow's demand, defined as the traffic rate it offers to a bottleneck, as a latent state to be inferred from network-side observations. When capacity is abundant, delivered throughput closely tracks demand. As load approaches capacity, however, throughput increasingly reflects available service, and under sustained saturation it bounds demand only from below. Inferring demand from throughput together with queue and loss signals separates offered traffic from service constraints, supporting congestion diagnosis, resource allocation, capacity planning, and scheduling-policy evaluation~\cite{medina2002traffic}. Existing data-driven approaches largely bypass latent demand and model observable traffic or QoS directly~\cite{perry2023dote}. Although effective when deployment conditions resemble those seen during training, such predictors often degrade under distribution shift~\cite{perry2023dote,tune2015spatiotemporal}, for example when link capacity changes or new traffic patterns emerge. Adapting to these changes typically requires new data collection and retraining, which can be costly or impractical in operational networks.

To address this limitation, we propose TG-DIN, a theory-guided demand inference network. TG-DIN combines a neural demand predictor with a differentiable theory layer that models a shared FIFO, tail-drop bottleneck through a window-level fluid approximation. The theory layer serves two roles. Run forward on sampled demand and network conditions, it generates mechanism-consistent training traces; placed after the predictor, it maps inferred demand, known capacity, and queue state to throughput, delay, and loss. Training minimizes the discrepancy between predicted and observed QoS, requiring neither demand labels nor collected traffic traces. Because capacity enters the layer explicitly as an input, a single model can operate across bottlenecks with different capacities within the generated operating range. The layer is also agnostic to how a flow's rate is produced, so changes induced by congestion control or application rate adaptation do not require retraining. Since such adaptation typically limits the duration of saturation, demand remains identifiable outside brief saturation periods.

We compare TG-DIN against GRU-LSTM and TFT-style direct predictors trained conventionally at a single bottleneck capacity. On synthetic traces, TG-DIN remains competitive near the baselines' training capacities and is substantially more robust under capacity shifts, achieving a per-flow throughput RMSE of 13.65~Mbps at 360~Mbps compared with 71.45--82.74~Mbps for the baselines. TG-DIN also outperforms baselines fine-tuned with up to 5\% target-capacity data. Although trained entirely on generated traces, it transfers to a controlled testbed without fine-tuning, including scenarios with a capacity drop from 120 to 10~Mbps and a transition from video streaming to a TCP bulk transfer. A complete training run requires approximately 36 minutes on a single GPU.

Our main contributions are threefold. First, we propose TG-DIN, a theory-guided framework that infers the demand each flow offers to a shared bottleneck from network-side observations without requiring demand supervision. Second, we introduce a differentiable theory layer that serves both as a randomized, mechanism-consistent trace generator and as a mapping from inferred demand to observable QoS, ensuring that the training data and learning objective are consistent by construction. Third, we evaluate TG-DIN on synthetic and real testbed traffic spanning capacity sweeps, traffic-pattern changes, dynamic capacity variations, and application changes, demonstrating strong robustness to capacity shifts without collected training data or deployment-specific fine-tuning.

\section{Related work}\label{sec:related_work}

\subsection{Network Measurement and QoS Prediction} 

Prior work has extensively studied the measurement, inference, and prediction of observable network performance metrics such as latency, throughput, and end-to-end path behavior. Systems such as iPlane~\cite{madhyastha2006iplane} and iPlaneNano~\cite{madhyastha2009iplane} demonstrated that large-scale path properties can be inferred from distributed measurements and lightweight probes. Other work modeled Internet delay spaces and latency distributions using compact statistical representations~\cite{zhang2006measurement}, while more recent studies have advanced latency and tail-latency estimation for network functions, datacenter fabrics, and large-scale services~\cite{iyer2022performance,zhao2023scalable,zhang2023latenseer}. A related line of research predicts application-level performance, including QoE for video streaming and QoS/QoE for real-time applications~\cite{balachandran2013developing,carofiglio2021characterizing}. These approaches model observable outcomes directly and treat QoS or QoE as the final prediction target. In contrast, our work focuses on latent demand inference. Rather than predicting network outcomes directly, TG-DIN estimates the per-flow demand offered to a shared bottleneck and relates it to observable throughput, delay, and loss through an explicit bottleneck model.

\subsection{Traffic Engineering under Demand Uncertainty}

Traffic engineering (TE) studies how to allocate network resources and select routes under changing traffic demands. Systems such as B4~\cite{jain2013b4} and SWAN~\cite{hong2013achieving} formulate TE as a centralized optimization problem driven by periodically estimated demand, largely separating demand estimation from control. More recent approaches incorporate machine learning to accelerate or replace parts of this pipeline. DOTE~\cite{perry2023dote} learns TE decisions directly from historical demand data, while TEAL~\cite{xu2023teal} and other neural WAN frameworks~\cite{alqiam2024transferable} improve optimization speed and generalization across network conditions. RedTE~\cite{gui2024redte} further targets real-time control at sub-second timescales. Unlike these systems, TG-DIN does not optimize routing or resource allocation. Instead, it addresses the complementary problem of inferring the latent demand offered to a shared bottleneck from observable QoS measurements and known service conditions.

\subsection{Hidden Traffic Demand Estimation}

Inferring unobserved traffic demand from indirect measurements has long been studied through traffic-matrix estimation, which reconstructs source--destination demand volumes from link measurements, routing information, or partial flow statistics~\cite{medina2002traffic,zhang2003information}. Recent work has incorporated machine learning and generative models to improve traffic-matrix recovery from incomplete or noisy observations~\cite{xu2021learning,yuan2023traffic,qiao2024autotomo}. These studies share the same underlying motivation as our work: observable measurements alone are often insufficient for effective network management. However, traffic-matrix estimation focuses on aggregate demand between network endpoints, typically in backbone and interdomain settings. TG-DIN instead estimates per-flow demand at a shared bottleneck from local observations of throughput, delay, and loss. Our objective is therefore not to reconstruct a network-wide traffic matrix, but to identify the demand of each flow that explains its observed QoS under shared service constraints.

\subsection{Deep Learning in Network Management}

Deep learning has become a prominent approach for network measurement, forecasting, and management. Representative examples include PERCEIVE~\cite{lee2020perceive}, which predicts cellular uplink throughput using a two-stage LSTM architecture; GRU-LSTM traffic predictors~\cite{shariff2025traffic}; and transformer-based frameworks such as the Temporal Fusion Transformer for traffic forecasting~\cite{kougioumtzidis2025mobile}. Recent surveys likewise identify recurrent and transformer architectures as the dominant deep-learning paradigms for network traffic prediction~\cite{aouedi2025deep}. Similar techniques have been applied to QoS forecasting through spatial-context-aware models and graph-based collaborative learning approaches~\cite{zhou2023spatial,hu2025gacl}. Across these studies, learned sequence models map historical measurements directly to future traffic or performance metrics. A major challenge for such models is distribution shift after deployment. Prior work has documented concept drift in operational networks~\cite{liu2023leaf}, while most drift-adaptation research in networking focuses on security applications such as intrusion detection, malware detection, and traffic classification~\cite{INSOMNIA,InvestigatingLabellessDriftAdaptation,FastLearning}. These approaches typically address changing environments through data collection and model updating. TG-DIN instead incorporates a bottleneck model into the learning process, using the same theory layer for both training-data generation and QoS-constrained demand inference, thereby reducing reliance on deployment-specific data and retraining.

\subsection{Differentiable Network Modeling}

Differentiable network models have primarily been developed to support gradient-based optimization of network control policies. dNE~\cite{ding2025learning} introduces a differentiable network evaluator that propagates gradients through traffic-engineering decisions, while differentiable queuing-network simulators~\cite{che2024differentiable} enable gradient-based optimization of scheduling and admission-control policies. TG-DIN uses differentiable modeling for a different purpose. Rather than optimizing a control policy, it infers the latent traffic demand that serves as input to a fixed bottleneck model. A demand predictor estimates latent demand, and gradients from a QoS reconstruction loss are propagated through the differentiable bottleneck model to train the predictor. The same model is also used to generate randomized training traces, ensuring consistency between the training data and the learning objective. This combination enables learning without demand supervision and improves robustness across operating conditions.

\section{Theory-Guided Demand Inference Network}\label{sec:method}

\subsection{Problem Formulation}\label{subsec:problem-formulation}

We consider multiple flows sharing a bottleneck whose capacity $C_t$ during measurement window $t$ is known. Let $N$ denote the number of flows, which may vary across traces. The available observations consist of per-flow throughput, queue-state features, and loss measurements. Let $X_t$ denote the recent observation history provided to the neural demand predictor $g_\phi$, including the current bottleneck capacity and queue state. Let $d_t \in \mathbb{R}_+^N$ denote the latent per-flow demand during window $t$. We define the demand of flow $i$ as the traffic rate it offers to the bottleneck, so $d_{i,t}\Delta t$ is the volume of new traffic contributed during a window of duration $\Delta t$. Traffic already present in the queue is represented separately by the backlog state. Demand is treated as exogenous within a window: it incorporates any endpoint congestion-control or application-level rate adaptation that occurred before window $t$, while feedback from the bottleneck to the sender during the window is not modeled. Consequently, the inferred demand represents the load offered after such adaptation rather than an application's unconstrained intent.

Because $d_t$ is not directly observable, our goal is to infer it from network-side measurements and use it to explain the resulting network behavior. We therefore seek a predictor 
\begin{equation} 
\hat d_t = g_\phi(X_t), 
\label{eq:demand-inference} 
\end{equation}
whose output is passed through a fixed forward model \begin{equation} 
\hat y_t = f_{\mathrm{theory}}(\hat d_t, C_t, s_t), 
\label{eq:theory-forward} 
\end{equation}
where $s_t$ contains the supplied backlog features, window duration, and buffer-regime parameter. The output $\hat y_t$ comprises predicted per-flow throughput together with delay and loss proxies. The predictor is trained by requiring $\hat y_t$ to match the observed outputs, without access to demand labels. The same forward model, run on sampled demand and network conditions, also generates the training traces (Section~\ref{subsec:Randomized}), ensuring that both the training data and the learning objective are derived from the same scheduling and queuing mechanism.

\begin{figure}[t]
    \centering
    \includegraphics[width=.99\linewidth]{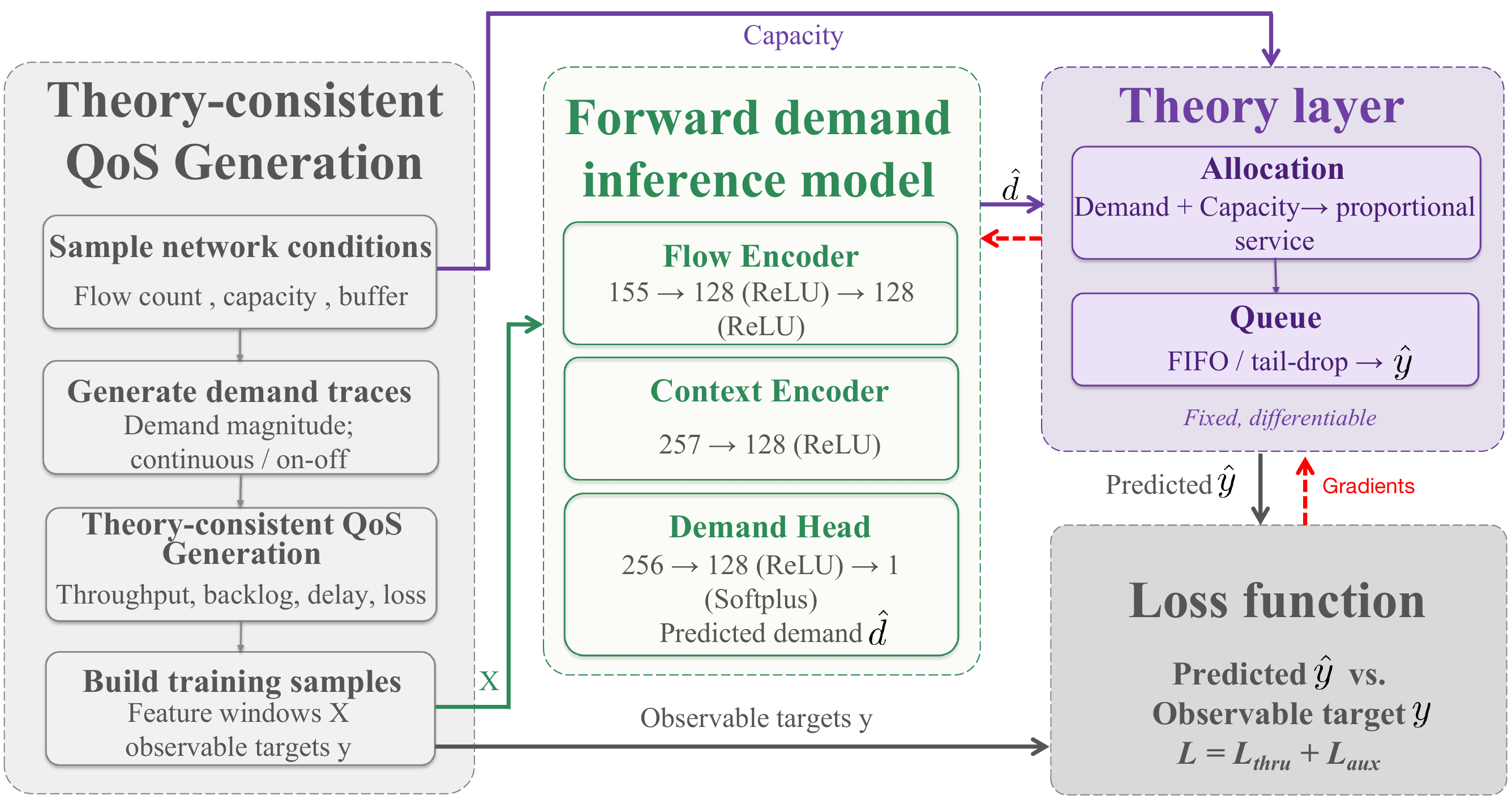}
    \caption{Overview of TG-DIN. The theory layer, run forward, generates the training traces; during training, it maps the predicted demand $\hat d$ and the capacity to observables $\hat y$. The loss compares $\hat y$ with the observed targets $y$, so demand is never supervised.}
    \label{fig:framework}
\end{figure}

This formulation also characterizes when demand is identifiable. When the bottleneck has spare capacity, delivered throughput closely tracks $d_{i,t}$. Under persistent saturation, observations constrain $d_{i,t}$ only from below because any demand beyond a flow's allocated share leaves no trace in its throughput. In practice, such periods tend to be brief because congestion control and rate adaptation reduce sending rates once the bottleneck becomes saturated, while backlog growth and loss provide additional information during these intervals. The throughput prediction produced by $f_{\mathrm{theory}}$ remains bounded by $C_t$ in all operating regimes. Fig.~\ref{fig:framework} illustrates the overall framework, from trace generation to learning through observable QoS.

\begin{figure*}[t]
    \centering
    \includegraphics[trim=1 25 0 28, clip, width=\textwidth]{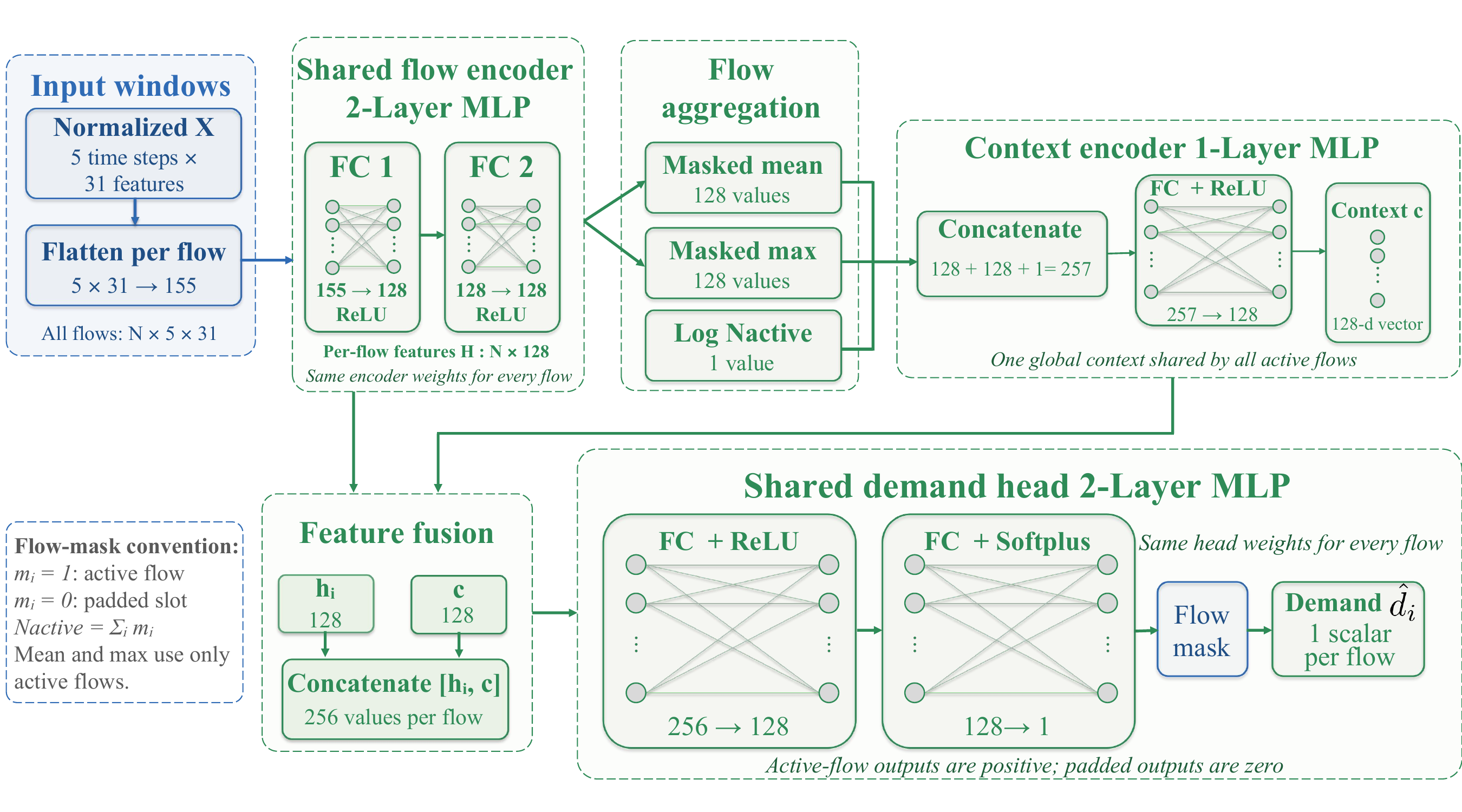}
    \caption{Demand inference model for one window. The same encoder is applied to every flow, the flow embeddings are pooled into one context vector shared by all flows, and the same head maps each flow embedding together with the context to one nonnegative demand. Padded slots are excluded from the pooling and their outputs are set to zero.}
    \label{fig:forward-model}
\end{figure*}

\subsection{Latent Demand Inference Model}
\label{subsec:latent-demand-model}

Fig.~\ref{fig:forward-model} depicts the latent demand inference model. As it shows, the latent demand predictor maps network-side observations to a nonnegative estimate of the rate each flow offers to the bottleneck. Because flows share a common bottleneck, each flow's observations reflect both its own behavior and its competition for service. We therefore implement $g_\phi$ as a set-based predictor with parameters shared across flows, allowing it to accommodate varying numbers of flows while combining per-flow temporal information with a shared bottleneck context.
For flow $i$, let $X_{i,t}\in\mathbb{R}^{K\times F}$ denote its feature window at time $t$, where $K$ is the history length and $F$ is the feature dimension. The window contains the throughput, queue-state, loss, and capacity features introduced in Section~\ref{subsec:problem-formulation}. A shared encoder maps the flattened window to a flow embedding:
\begin{equation}
    h_{i,t}
    =
    e_\phi\!\left(\operatorname{vec}(X_{i,t})\right),
    \label{eq:flow-encoder}
\end{equation}
where $\operatorname{vec}(\cdot)$ preserves the temporal ordering within the window. The encoder $e_\phi$ consists of two fully connected layers with ReLU activations and is applied identically to all flows.

Because traces contain varying numbers of flows, inputs are padded to a fixed size and accompanied by a binary mask indicating active flows. To capture the shared bottleneck state, the embeddings of all active flows are aggregated using mean and element-wise maximum pooling. These summaries are combined with the logarithm of the number of active flows to form a shared context:
\begin{equation}
    c_t
    =
    \rho_\phi\!\left(
        \left[
        \operatorname{mean}_{j} h_{j,t},
        \operatorname{max}_{j} h_{j,t},
        \log N
        \right]
    \right),
    \label{eq:bottleneck-context}
\end{equation}
where pooling operates over the active flows, $[\cdot]$ denotes concatenation, and $\rho_\phi$ is a fully connected layer followed by a ReLU activation. Mean pooling summarizes collective flow behavior, max pooling preserves dominant embedding responses, and $\log N$ captures the degree of contention at the bottleneck.

A shared demand head combines each flow embedding with the bottleneck context to produce
\begin{equation}
    \hat d_{i,t}
    =
    \operatorname{softplus}\!\left(
        \psi_\phi([h_{i,t},c_t])
    \right),
    \label{eq:shared-demand-head}
\end{equation}
where $\psi_\phi$ consists of a fully connected hidden layer with a ReLU activation followed by a scalar output layer. The demand estimate for padded slots is set to zero. The Softplus activation enforces nonnegative outputs while allowing estimated demand to exceed bottleneck capacity when appropriate. Each demand estimate depends on both the flow's own observation history and the shared bottleneck context. Because the encoder and demand head are shared across flows and the aggregation operators are permutation-invariant, reordering flows produces the same reordering of outputs, making the model independent of flow ordering.
The resulting demand vector $\hat d_t$ serves as the latent input to the scheduling and queueing layer described next. During training, gradients from the QoS reconstruction loss propagate through that layer to update the demand predictor, eliminating the need for demand supervision. As discussed in Section~\ref{subsec:problem-formulation}, throughput observations strongly constrain demand when spare capacity is available, while under sustained saturation the demand exceeding a flow's allocated share is constrained primarily through the backlog and loss outputs.

\subsection{Theory Layer}
\label{subsec:theory-layer}

The theory layer connects inferred latent demand to observable QoS through an explicit model of bottleneck service and queuing. As a proof of concept, we consider a shared bottleneck with FIFO service and tail-drop buffer management. FIFO serves packets in their arrival order~\cite{cisco-cong-mgmt}, and tail-drop discards newly arriving packets when the queue is full~\cite{rfc7567}. We approximate these mechanisms at the level of measurement windows with a fluid model of traffic rates and backlog volumes: proportional allocation represents per-flow service, aggregate backlog and service rate determine queuing delay, and a per-flow buffer-overflow approximation determines loss.

Consider $N$ active flows sharing a bottleneck of capacity $C$ over a measurement window of duration $\Delta t$. For readability, we omit the time index $t$ below. Let $\hat d_i$ denote the inferred demand of flow $i$, that is, the rate of new traffic it offers during the window, and let $B_i$ denote its backlog at the start of the window. The backlog is supplied as part of the queue state $s$ rather than propagated by the layer from one window to the next.

\textbf{Scheduling Module:}
The scheduling module assigns a service rate to each flow from its inferred demand, its backlog, and the available capacity. The total work of flow $i$ during the window is
\begin{equation}
    q_i = B_i + \hat d_i \Delta t.
    \label{eq:available-work}
\end{equation}
We define the effective workload rate $w_i$ as the service rate needed to clear this work within the window:
\begin{equation}
    w_i = \frac{q_i}{\Delta t}
        = \hat d_i + \frac{B_i}{\Delta t}.
    \label{eq:eff-demand}
\end{equation}
Here, $\hat d_i$ represents newly offered traffic, whereas $w_i$ also accounts for traffic already awaiting service.
Service is allocated by a proportional, work-conserving rule:
\begin{equation}
    \hat a_i =
    \begin{cases}
        w_i,
        & \displaystyle \sum_{j=1}^{N} w_j \le C, \\
        C \dfrac{w_i}{\sum_{j=1}^{N} w_j},
        & \text{otherwise}.
    \end{cases}
    \label{eq:alloc}
\end{equation}
When capacity is sufficient, all work is served within the window. Otherwise, the capacity is divided in proportion to the effective workload rates. The allocation satisfies
\begin{equation}
    0 \le \hat a_i \le w_i,
    \qquad
    \sum_{i=1}^{N} \hat a_i
    = \min\!\left(C,\sum_{i=1}^{N} w_i\right).
    \label{eq:service-feasibility}
\end{equation}
Classical fluid service models, e.g., Generalized Processor Sharing (GPS)~\cite{parekh1993gps}, provide a tractable abstraction of capacity sharing among concurrent flows. Unlike GPS, which assigns each flow a fixed weight, Eq.~\eqref{eq:alloc} makes each flow's share proportional to its workload. This approximates a FIFO bottleneck, where a flow's share of service follows its share of the traffic awaiting service, without tracking individual packet ordering. Under saturation, the allocation depends only on the relative workload rates $w_i/\sum_j w_j$, so throughput constrains the relative demand of the flows, while the backlog and loss carry the remaining information about its absolute level.

\textbf{Queuing and Loss Module:}
Given the service allocation, the queuing and loss module produces the observable QoS estimates: throughput, delay, and packet loss. The volume transmitted during the window is bounded by the available work and the allocated service,
\begin{equation}
    \mathrm{sent}_i
    = \min\!\left(q_i,\hat a_i\Delta t\right),
    \label{eq:sent}
\end{equation}
and the resulting throughput estimate is
\begin{equation}
    \hat r_i = \frac{\mathrm{sent}_i}{\Delta t}.
    \label{eq:throughput}
\end{equation}
By Eqs.~\eqref{eq:alloc} and~\eqref{eq:service-feasibility}, $\hat r_i \le C$ for every flow and $\sum_i \hat r_i \le C$, so throughput predictions never exceed the bottleneck capacity.
To represent the queuing condition shared by all flows at the bottleneck, we compute a delay proxy from the aggregate backlog and the aggregate allocated service rate:
\begin{equation}
    \hat\tau_i
    = \min\!\left(
        \tau_{\max},
        \frac{\sum_{j=1}^{N} B_j}
        {\max\!\left(\sum_{j=1}^{N}\hat a_j,A_{\min}\right)}
      \right),
    \label{eq:delay}
\end{equation}
where $A_{\min}>0$ is a service-rate floor and $\tau_{\max}$ is the largest delay the proxy represents. Every active flow therefore receives the same bottleneck delay estimate.
The work remaining after service is
\begin{equation}
    q_i^{\mathrm{rem}}
    = q_i-\mathrm{sent}_i.
    \label{eq:remaining-work}
\end{equation}
We approximate loss by applying the same effective buffer threshold $B_{\max}=C\beta$ to each flow separately, where $\beta$ is a buffer-duration parameter. This per-flow threshold simplifies the shared tail-drop buffer while retaining its dependence on capacity. The overflow volume and loss proxy are
\begin{equation}
    \mathrm{drop}_i
    = \max\!\left(q_i^{\mathrm{rem}}-B_{\max},0\right),
    \hat\ell_i
    = \min\!\left(
        1,\frac{\mathrm{drop}_i}{q_i}
       \right).
    \label{eq:loss}
\end{equation}
Together, Eqs.~\eqref{eq:available-work}-\eqref{eq:loss} define a fixed forward map from inferred demand, capacity, and queue state to observable QoS estimates. The theory layer has no trainable parameters and is differentiable almost everywhere, so losses on the predicted throughput and, where available, on delay and loss propagate back to the demand predictor. Latent demand is therefore learned without demand values as supervision, which addresses the absence of demand labels in the measurement setting considered here.


\subsection{Theory-Guided Model Implementation}\label{subsec:Randomized}

TG-DIN uses the theory layer in two roles. Run forward on sampled demand and network conditions, it generates observable traffic traces; placed after the demand predictor, it maps demand estimates back to QoS during training. Because both roles use the same scheduling and queuing mechanism, the generated observations are consistent with the forward model used for learning, and the predictor learns from them without demand supervision.

\textbf{Randomized Regime Generation:}
For each trace, we sample the number of concurrent flows, a slowly varying capacity trajectory within a sampled capacity band, the demand magnitude and temporal pattern of each flow, and a buffer-duration parameter from one of several buffer regimes. Each flow follows either a \emph{continuous} pattern, representing sustained traffic such as WebRTC calls, or an \emph{on-off} pattern, representing bursty transfers such as live video streaming, which alternates bursts with multi-second idle intervals~\cite{carrascosa2022cloud}. The durations and rates of these states are randomized, so the generator covers a broad family of temporal behaviors rather than reproducing specific applications. Demand is sampled independently of capacity, consistent with the exogenous-demand assumption of Section~\ref{subsec:problem-formulation}. The generator applies the theory layer window by window, carrying the backlog remaining after service and overflow into the next window, and records the resulting throughput, delay, and loss. Observation histories and their QoS targets are then extracted from these traces.
Traces are generated on demand and discarded after each training round, then replaced with new samples. This procedure requires no collected, stored, or fixed training corpus, and training coverage can be extended by changing the generation ranges rather than by collecting data for each new condition. The ranges used in our experiments are given in Section~\ref{subsec:setup}.

\textbf{Training Objective:}
Although the generator knows the sampled demand, these values are never given to the predictor or used as targets, so demand is learned without ground truth, as it must be in real deployments. The theory layer maps the predictor's output $\hat d_t$ to per-flow throughput $\hat r_t$, delay $\hat\tau_t$, and loss $\hat\ell_t$, which are compared with the observed $r_t$, $\tau_t$, and $\ell_t$. Throughput provides the primary supervision. To handle its wide dynamic range, we compare it mainly in log space, with a small linear-space term that preserves calibration in absolute scale:
\begin{equation}
\mathcal{L}_{\mathrm{thr}}
=
\mathrm{MSE}\!\left(\log(1+\hat r_t),\log(1+r_t)\right)
+
\lambda_{\mathrm{lin}}\,\mathrm{SmoothL1}(\hat r_t, r_t).
\end{equation}
The log-space term keeps high-throughput samples from dominating and gives low-throughput samples a meaningful gradient. 
Loss adds supervision on buffer overflow and is the only output of the theory layer that can depend on the absolute demand level once the bottleneck saturates. Delay adds supervision below saturation, since under saturation $\sum_j \hat a_j = C$ and Eq.~\eqref{eq:delay} no longer depends on $\hat d$. The auxiliary loss is:
\begin{equation}
\mathcal{L}_{\mathrm{aux}}
=
\lambda_{\tau}\,\mathrm{MSE}(\hat\tau_t,\tau_t)
+
\lambda_{\ell}\,\mathrm{MSE}(\hat\ell_t,\ell_t).
\label{eq:auxiliary-qos-loss}
\end{equation}
The overall objective, computed over active flows only, is
\begin{equation}
\mathcal{L} = \mathcal{L}_{\mathrm{thr}} + \mathcal{L}_{\mathrm{aux}}.
\label{eq:training-objective}
\end{equation}
Gradients of $\mathcal{L}$ propagate through the fixed theory layer to the demand predictor, which therefore learns demand estimates consistent with the observed QoS across the generated operating conditions.

\section{Evaluation}\label{sec:evaluation}

\subsection{Experimental Setup}\label{subsec:setup}

\noindent\textbf{Platform:}
All experiments run on a single Linux workstation (Ubuntu 22.04 LTS, kernel 6.8) with an NVIDIA GeForce RTX 5090 GPU and 64 GB of memory, using PyTorch 2.4 and Python 3.10.

\noindent\textbf{Training Configuration:}
We instantiate the randomized generation procedure of Section~\ref{subsec:Randomized} with the settings in Table~\ref{tab:exp-setup}. Each refresh round contains 200 new traces, each with two to six flows sharing one bottleneck, and traces are refreshed every 30 epochs. Capacities follow slowly varying trajectories within eight bands covering $[3,600]$~Mbps. Per-flow demand magnitudes span $[0.05,600]$~Mbps across four overlapping regimes: \emph{small} (up to 5~Mbps), \emph{light} (about 1--8~Mbps), \emph{medium} (about 3--20~Mbps), and \emph{heavy} (10~Mbps and above). Traces are divided into 0.2~s measurement windows, and the predictor uses a history of $K=5$ windows.

\noindent\textbf{Optimization and Model Selection:}
TG-DIN is trained with Adam (Table~\ref{tab:exp-setup}). Within each round, traces are split at the trace level into training, validation, and held-out subsets. Because the training traces change across rounds, checkpoint selection and early stopping use a fixed global validation set, generated once before training from four traces per capacity band. As in Section~\ref{subsec:Randomized}, the demand values used by the generator are never used as supervision.

\begin{table}[ht!]
\centering
\caption{Training configuration and key hyperparameters for TG-DIN and the baselines.}
\label{tab:exp-setup}
\footnotesize
\setlength{\tabcolsep}{4pt}
\begin{tabular}{@{}lll@{}}
\toprule
\textbf{Parameter} & \multicolumn{2}{l}{\textbf{Value}} \\
\midrule
\multicolumn{3}{@{}l}{\textit{Randomized training data}} \\
Observation window $K$ & \multicolumn{2}{l}{5 windows (0.2\,s each)} \\
Traces per refresh round & \multicolumn{2}{l}{200, refreshed every 30 epochs} \\
Flows per trace & \multicolumn{2}{l}{2--6} \\
Capacity range & \multicolumn{2}{l}{$[3, 600]$ Mbps (8 bands)} \\
Demand magnitude range & \multicolumn{2}{l}{$[0.05, 600]$ Mbps} \\
Demand regimes & \multicolumn{2}{l}{Small, light, medium, heavy} \\
Traffic patterns & \multicolumn{2}{l}{Continuous, on-off} \\
Mean ON / OFF duration & \multicolumn{2}{l}{1--4\,s / 1--6\,s} \\
\midrule
 & \textit{TG-DIN} & \textit{Baselines} \\
\cmidrule(lr){2-2}\cmidrule(l){3-3}
Optimizer & Adam & Adam \\
Learning rate & $10^{-4}$ & $10^{-3}$ \\
LR scheduler & -- & ReduceLROnPlateau \\
\quad decay / patience & -- & $\times 0.5$ / 3 \\
Weight decay & $10^{-5}$ & -- \\
Batch size & 256 & 256 \\
Gradient clipping ($\ell_2$) & -- & 5.0 \\
Max epochs / early stop & 60 / 10 & 60 / 10 \\
Data split & 80/10/10\,\% per round & 90/10\,\% \\
Global validation set & 4 traces per band & -- \\
\bottomrule
\end{tabular}
\end{table}

\subsection{Baselines}

We compare TG-DIN against direct predictors that map the same causal observation window ($K=5$) and the same features directly to per-flow throughput, without inferring latent demand or using a theory layer:
\begin{itemize}
\item \textbf{Direct-GRU-LSTM~\cite{shariff2025traffic}} represents recurrent direct prediction: a stacked GRU and LSTM followed by a two-layer MLP head with a softplus output.
\item \textbf{Direct-TFT-style~\cite{kougioumtzidis2025mobile}} represents attention-based direct prediction: a linear input projection, an LSTM encoder, a multi-head self-attention block, and a feed-forward head with a softplus output. It keeps the recurrent and attention components of the Temporal Fusion Transformer that are relevant to our task and omits the rest.
\end{itemize}

\textbf{Baseline Training:}
The baselines use the same log-based normalization and throughput loss as TG-DIN, namely log-space MSE plus $0.01\cdot\mathrm{SmoothL1}$ in the original scale. They have no delay or loss terms because they output throughput only. Following the conventional practice of training a direct predictor on data from the condition under which it was collected, each baseline is trained on 120 synthetic traces at one fixed bottleneck capacity, with separate models at 20, 40, and 60~Mbps, and is then evaluated across a wider range of capacities. Remaining hyperparameters are listed in Table~\ref{tab:exp-setup}.

\textbf{Target-Adapted Fine-Tuning:}
Updating a deployed model with newly collected samples is the common response to distribution shift in networking~\cite{INSOMNIA, InvestigatingLabellessDriftAdaptation, FastLearning}. We instantiate this step by fine-tuning both baselines on calibration sets drawn from the target capacity, which stand in for newly collected target traces. We consider budgets of 1\% and 5\% of the target windows and two strategies: \emph{full} fine-tuning of all parameters and \emph{last-layer} fine-tuning of the output head only. TG-DIN is evaluated without any fine-tuning.


\subsection{Test Setting}

\textbf{Synthetic Test Grid:}
The primary synthetic benchmark covers seven bottleneck capacities, $\{20,40,60,120,200,280,360\}$~Mbps. At each capacity, we generate 40 two-flow test traces of 400 windows each, with fixed capacity within a trace and demand levels and traffic patterns sampled by the training generation rules. Test seeds are disjoint from those used for training and model selection.
We also construct controlled drift traces with the same generator. In each drift trace, capacity is fixed, $u_1$ is a continuous flow of about 4~Mbps, and the traffic of $u_0$ changes once, at window 200 of 400. The three scenarios in Table~\ref{tab:scenario-def} change the demand level of $u_0$, its temporal pattern, or both. Each scenario is generated at 20, 60, 280, and 360~Mbps with 20 traces per capacity, and the same demand traces are replayed at every capacity so that the effect of capacity is isolated from the change in traffic. The 20 and 60~Mbps cases are capacity-constrained, whereas the 280 and 360~Mbps cases remain uncongested.
These benchmarks support two evaluations: cross-capacity generalization without adaptation (Section~\ref{subsec:cross-capacity-no-adap}) and comparison with target-adapted fine-tuning (Section~\ref{subsec:adap}).

\begin{table}[ht!]
\centering
\caption{Drift scenarios: traffic of $u_0$ before and after the change (rates in Mbps; $\mu$ is the mean demand).}
\label{tab:scenario-def}
\footnotesize
\setlength{\tabcolsep}{2pt}
\renewcommand{\arraystretch}{1.1}
\begin{tabular}{>{\raggedright\arraybackslash}p{0.24\columnwidth}>{\raggedright\arraybackslash}p{0.27\columnwidth}>{\raggedright\arraybackslash}p{0.25\columnwidth}>{\raggedright\arraybackslash}p{0.11\columnwidth}}
\toprule
\textbf{Scenario} &
\textbf{$u_0$ before} &
\textbf{$u_0$ after} &
\textbf{Change} \\
\midrule
demandOnly &
on-off, 42-78, $\mu\!\approx\!16.2$ &
on-off, 84-156, $\mu\!\approx\!35.0$ &
level \\
patternOnly &
on-off, 42-78, $\mu\!\approx\!16.1$ &
cont., $\mu\!\approx\!16.2$ &
pattern \\
patternDemand &
on-off, 42-78, $\mu\!\approx\!16.6$ &
cont., $\mu\!\approx\!33.9$ &
both \\
\bottomrule
\end{tabular}
\vspace{-2mm}
\end{table}

\begin{figure*}[hb]
    \centering
    \begin{subfigure}[t]{0.32\linewidth}
        \centering
        \includegraphics[width=\linewidth]{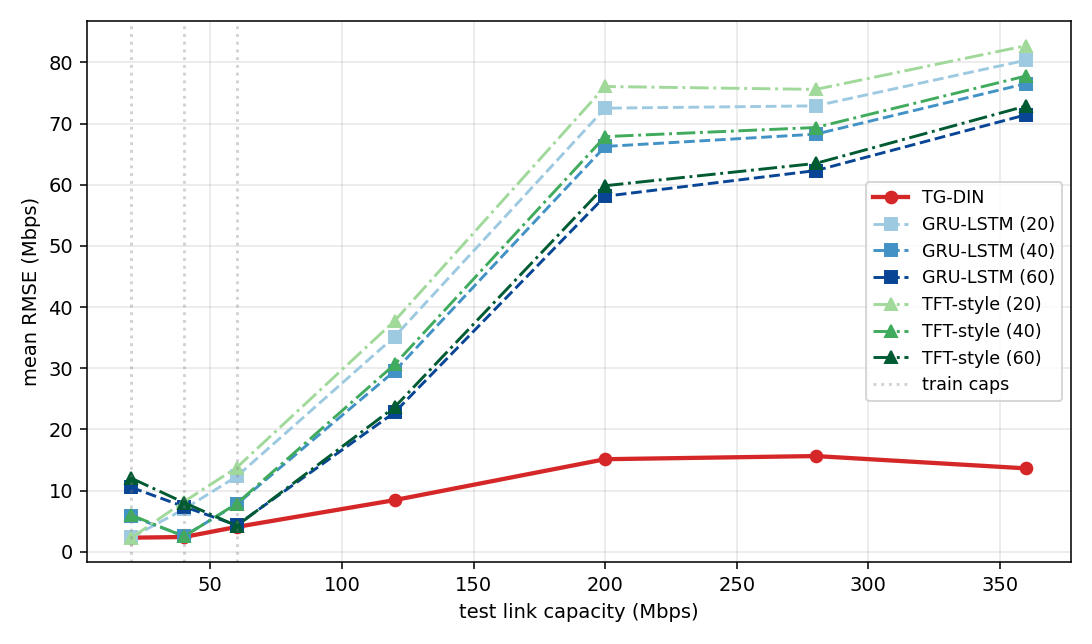}
        \caption{Mean RMSE}
    \end{subfigure}
     \hfill
    \begin{subfigure}[t]{0.32\linewidth}
        \centering
        \includegraphics[width=\linewidth]{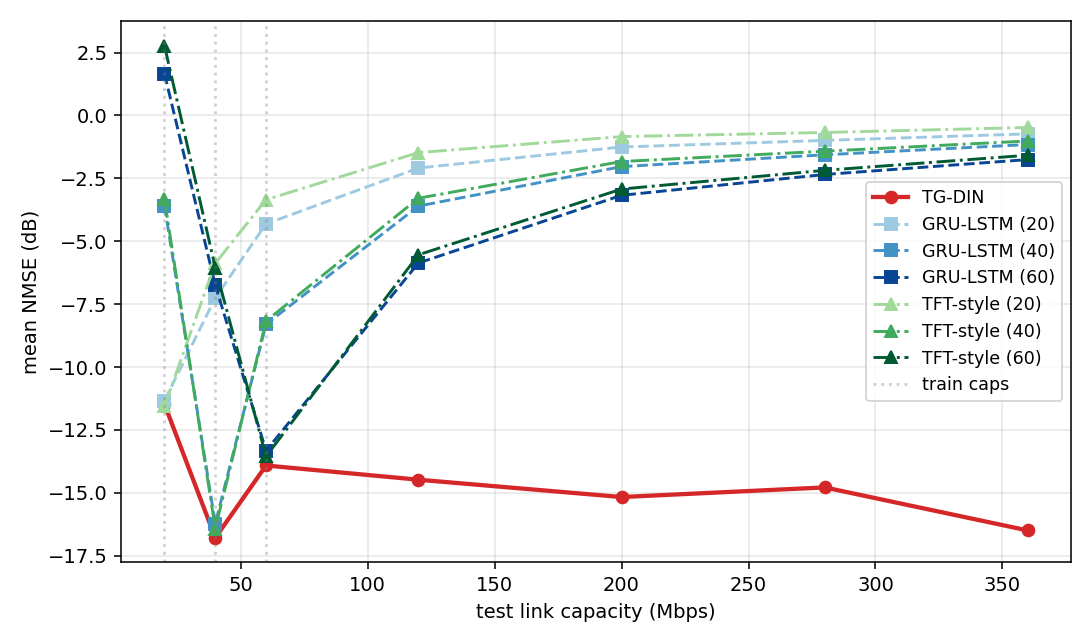}
        \caption{Mean NMSE (dB)}
    \end{subfigure} 
     \hfill
    \begin{subfigure}[t]{0.32\linewidth}
        \centering
        \includegraphics[width=\linewidth]{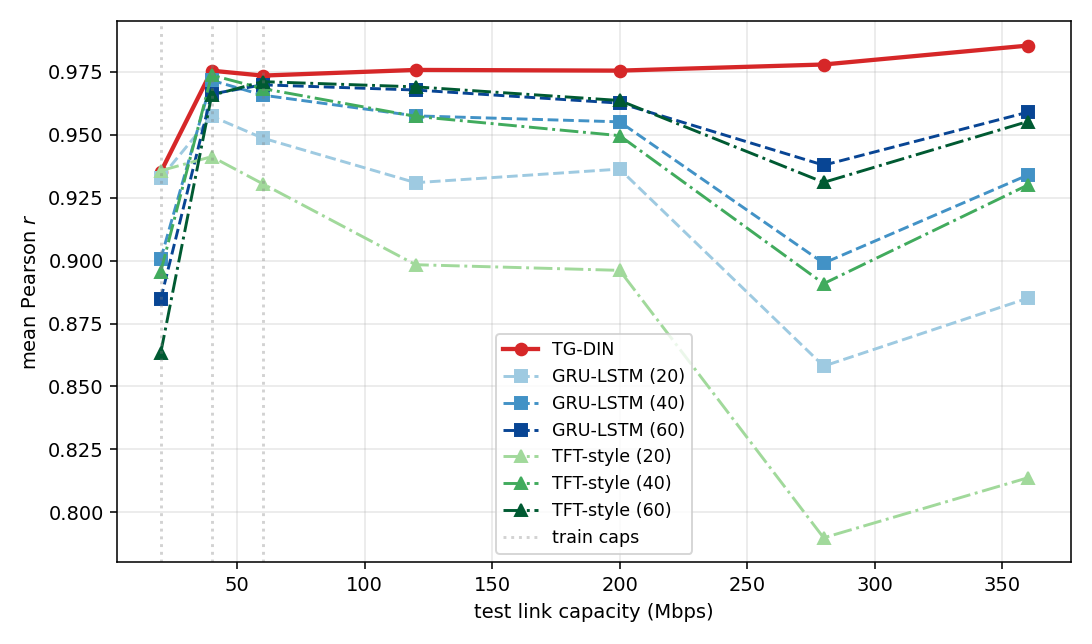}
        \caption{Mean Pearson $r$}
    \end{subfigure}
    \caption{Cross-capacity synthetic transfer results without adaptation. }
    \label{fig:cross-capacity-main}
\end{figure*}

\textbf{Real-Traffic Captures: }
To evaluate whether the proposed framework transfers beyond the synthetic setting, we conduct a preliminary real‑traffic experiment in a controlled network environment. A laptop running Linux acts as the shared gateway and applies the bottleneck rate on the outbound path with the Linux traffic control framework (\texttt{tc}), and one client laptops running two flows (emulating two users) reach it over Wi-Fi, so both flows compete over the same rate-limited link. This setup creates controlled contention between the two users and induces observable interactions in their achieved throughput. One flow ($u_0$) streams YouTube video and the other ($u_1$) runs a real-time video conferencing application (FaceTime). Because the bottleneck rate is configured at the gateway, the capacity of each capture is known by construction, and the gateway can also sample the queue state of the shaper in parallel with the packet capture. Packets are aggregated into 0.2\,s windows with the same pipeline used for the synthetic traces, so the synthetic and the real evaluations share one input interface. The individual captures and the conditions they cover are described together with the corresponding results.

\begin{figure*}[ht!]
    \centering
    \begin{subfigure}[t]{0.24\textwidth}
        \centering
        \includegraphics[width=\linewidth]{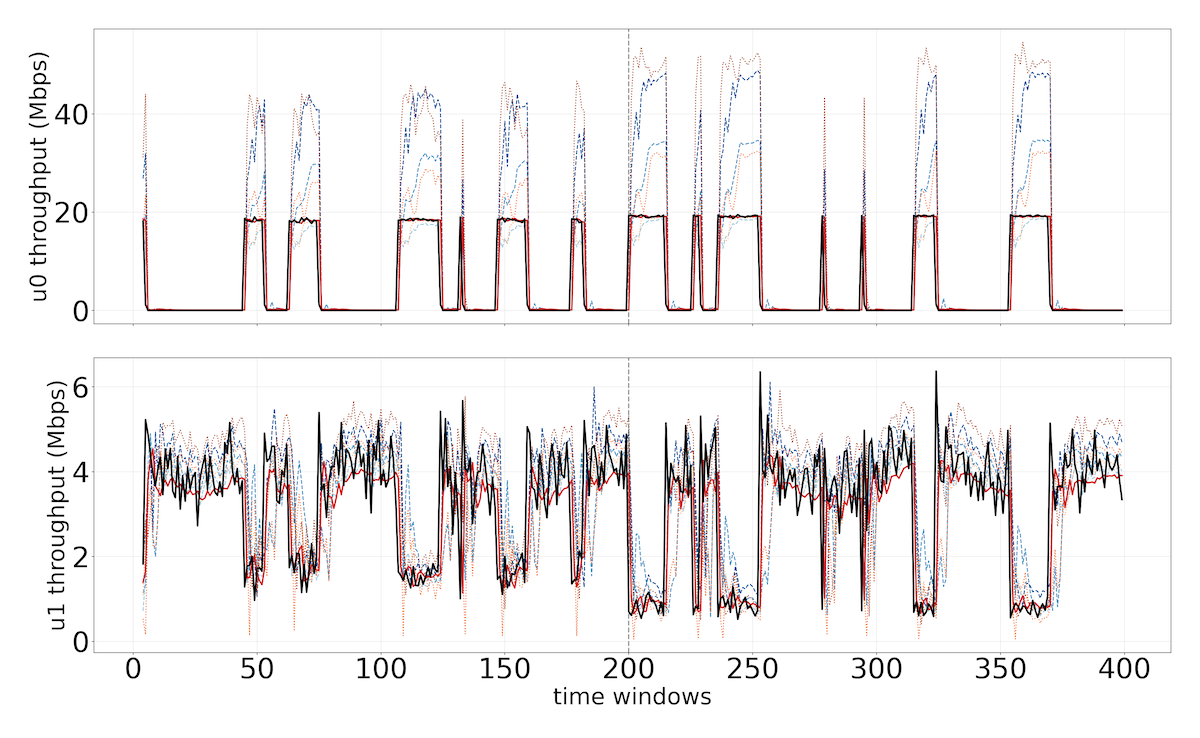}
        \caption{\textit{demandOnly}, $C=20$ Mbps}
    \end{subfigure}
    \hfill
    \begin{subfigure}[t]{0.24\textwidth}
        \centering
        \includegraphics[width=\linewidth]{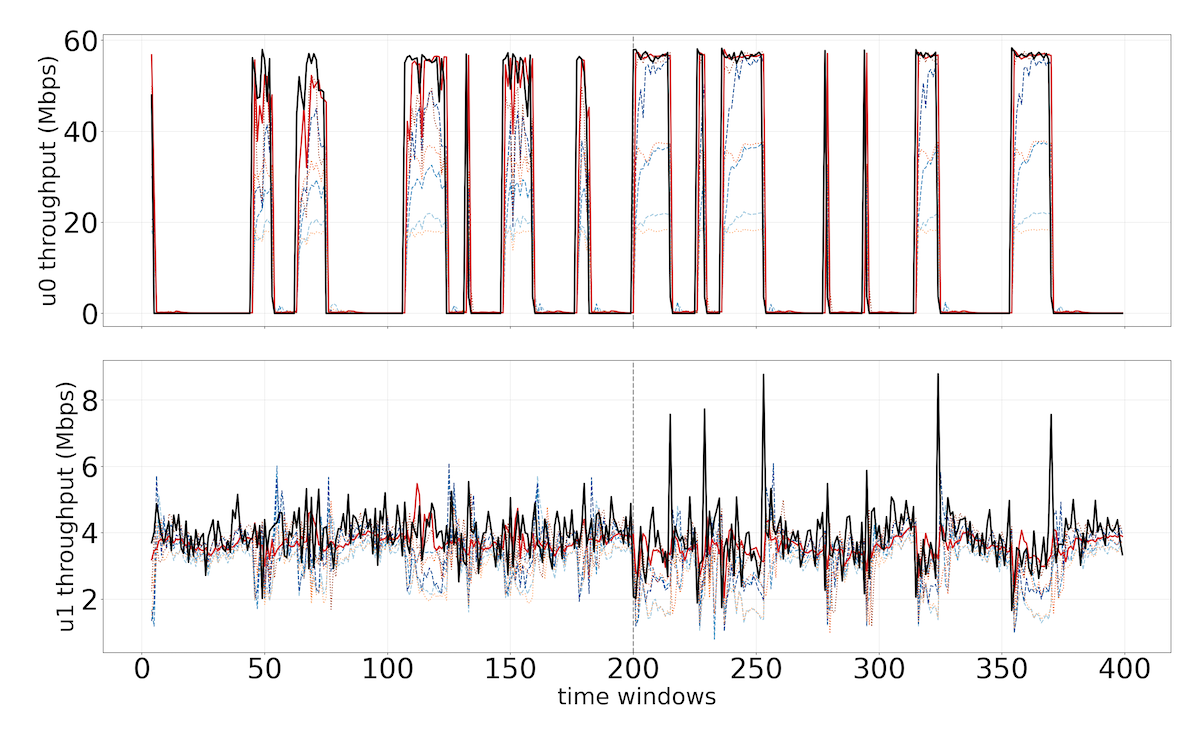}
        \caption{\textit{demandOnly}, $C=60$ Mbps}
    \end{subfigure}
    \hfill
    \begin{subfigure}[t]{0.24\textwidth}
        \centering
        \includegraphics[width=\linewidth]{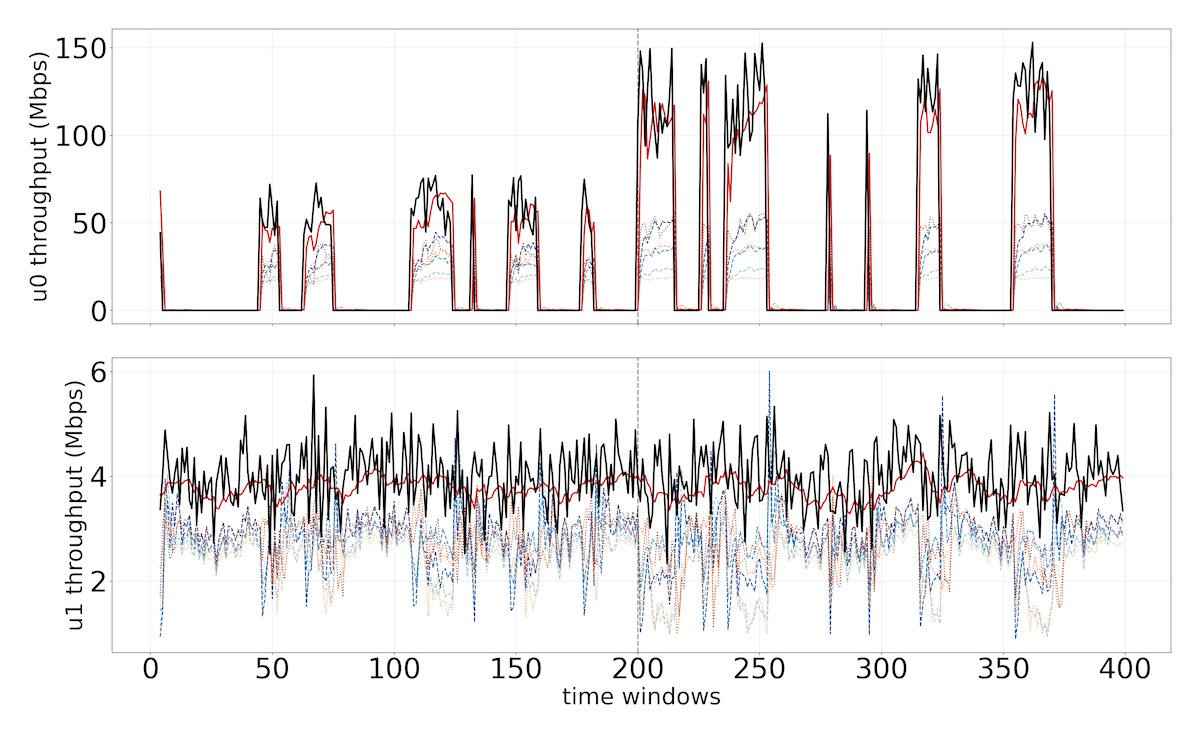}
        \caption{\textit{demandOnly}, $C=280$ Mbps}
    \end{subfigure}
    \hfill
    \begin{subfigure}[t]{0.24\textwidth}
        \centering
        \includegraphics[width=\linewidth]{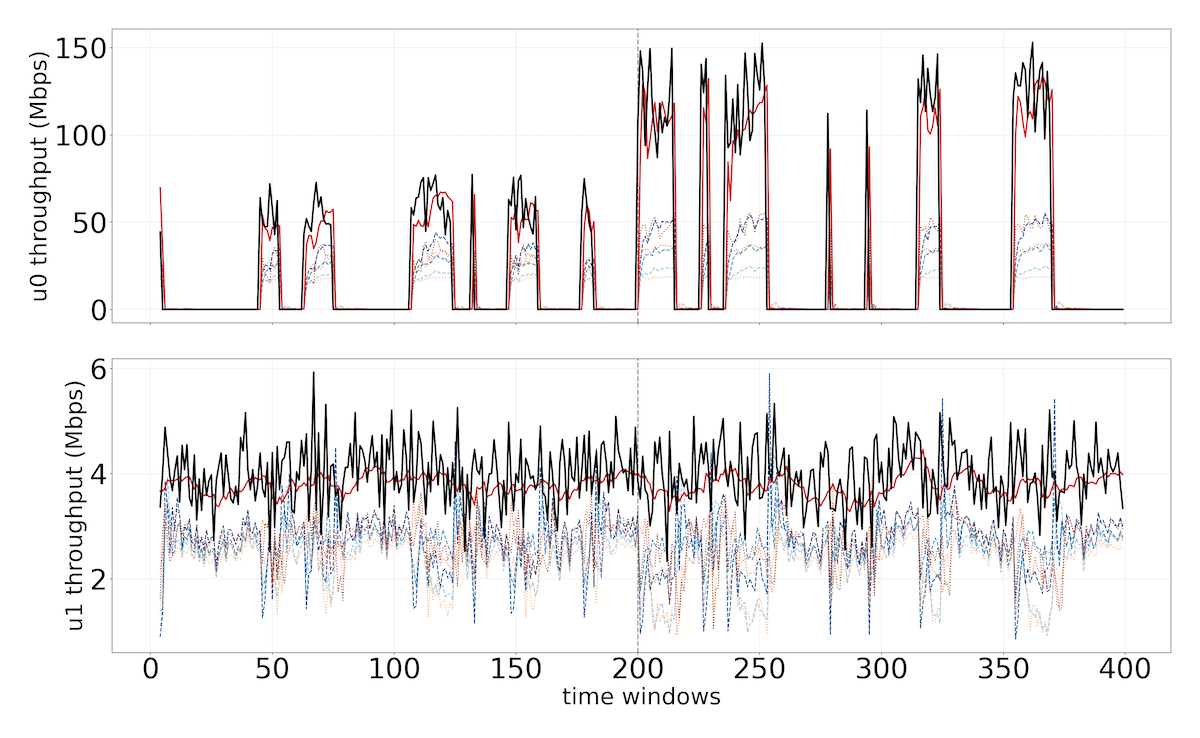}
        \caption{\textit{demandOnly}, $C=360$ Mbps}
    \end{subfigure}

    \vspace{0.5em}

    \begin{subfigure}[t]{0.24\textwidth}
        \centering
        \includegraphics[width=\linewidth]{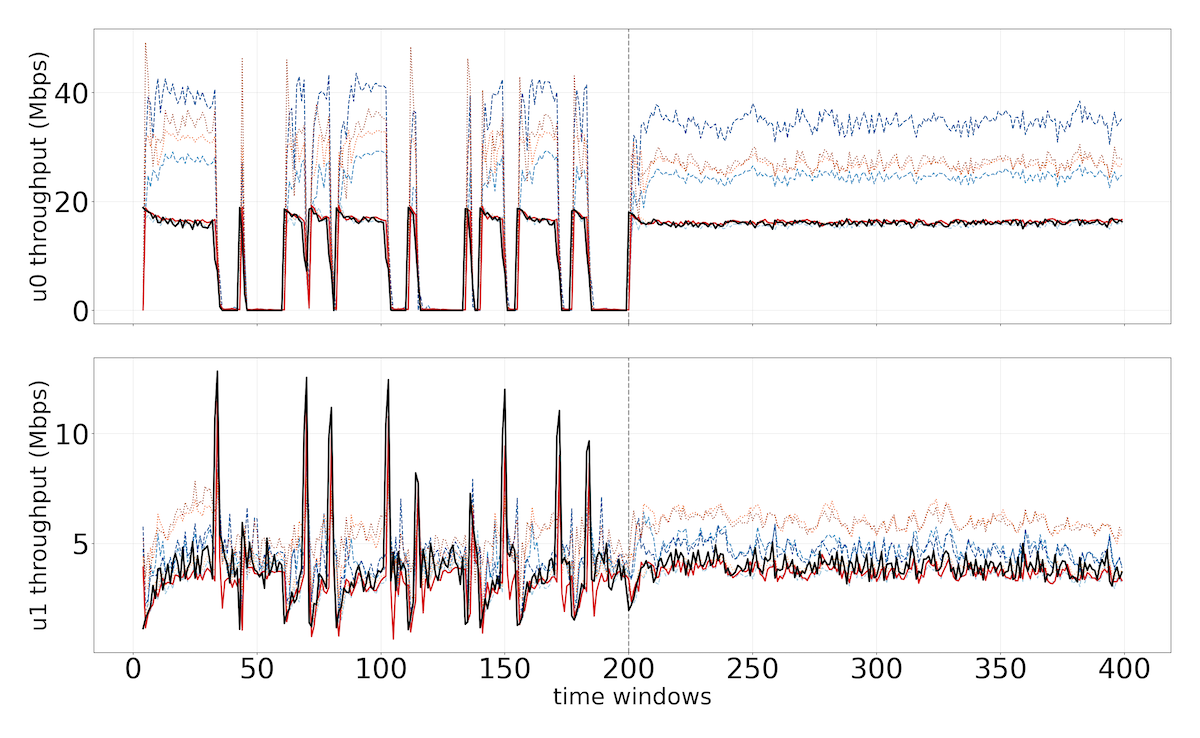}
        \caption{\textit{patternOnly}, $C=20$ Mbps}
    \end{subfigure}
    \hfill
    \begin{subfigure}[t]{0.24\textwidth}
        \centering
        \includegraphics[width=\linewidth]{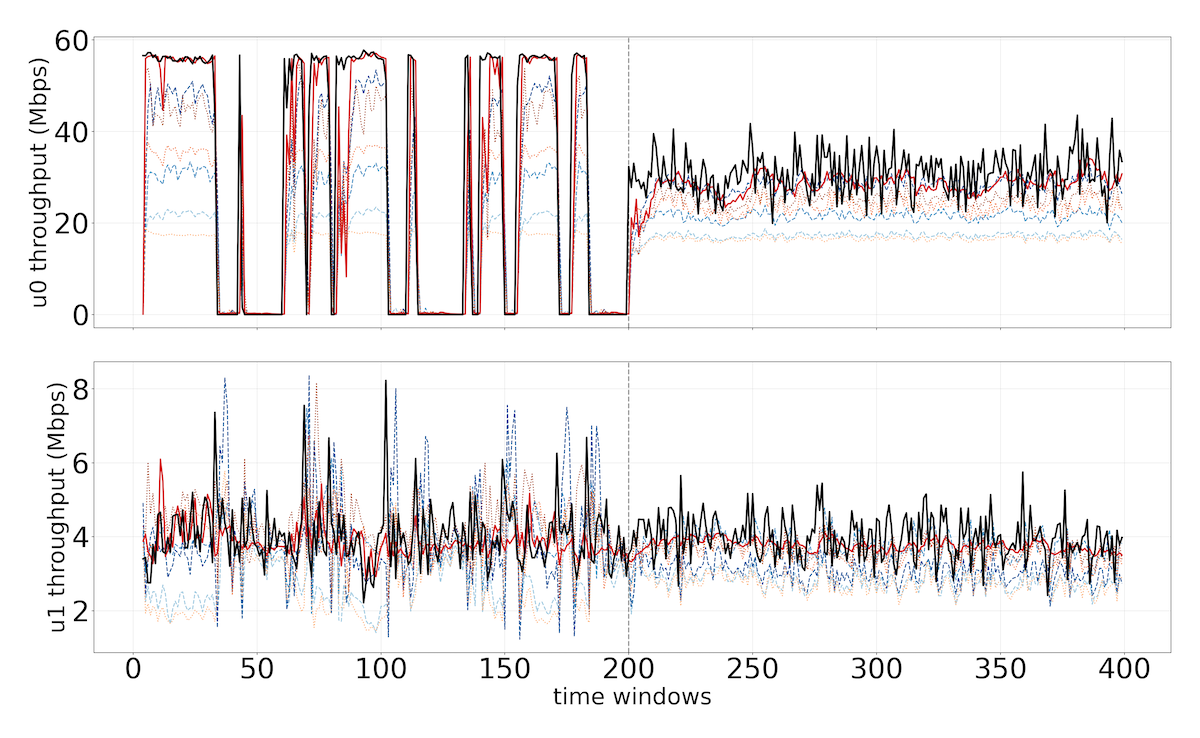}
        \caption{\textit{patternOnly}, $C=60$ Mbps}
    \end{subfigure}
    \hfill
    \begin{subfigure}[t]{0.24\textwidth}
        \centering
        \includegraphics[width=\linewidth]{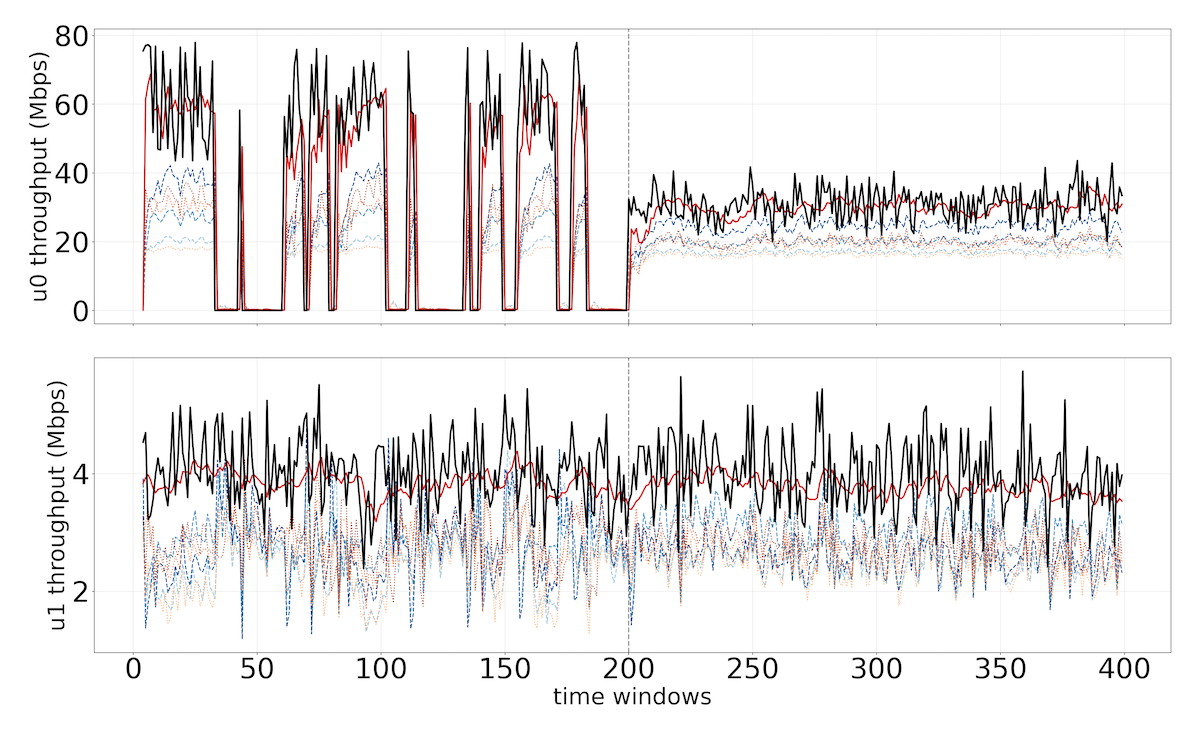}
        \caption{\textit{patternOnly}, $C=280$ Mbps}
    \end{subfigure}
    \hfill
    \begin{subfigure}[t]{0.24\textwidth}
        \centering
        \includegraphics[width=\linewidth]{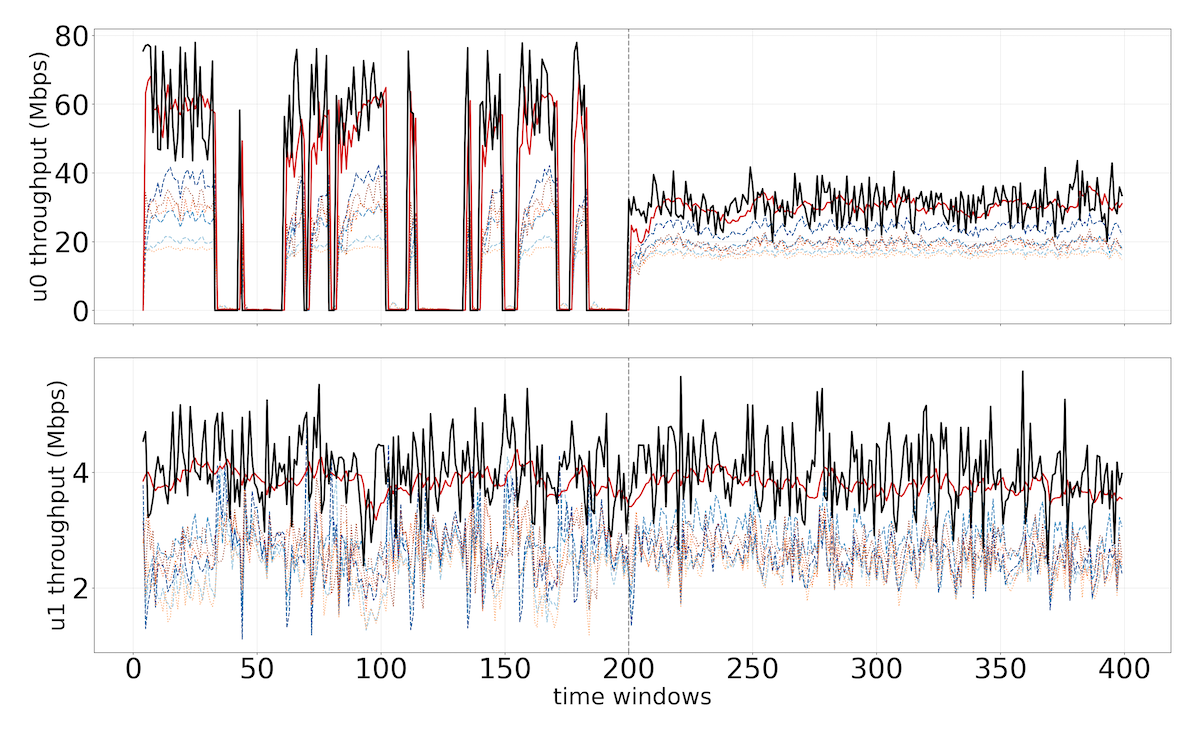}
        \caption{\textit{patternOnly}, $C=360$ Mbps}
    \end{subfigure}

    \vspace{0.5em}

    \begin{subfigure}[t]{0.24\textwidth}
        \centering
        \includegraphics[width=\linewidth]{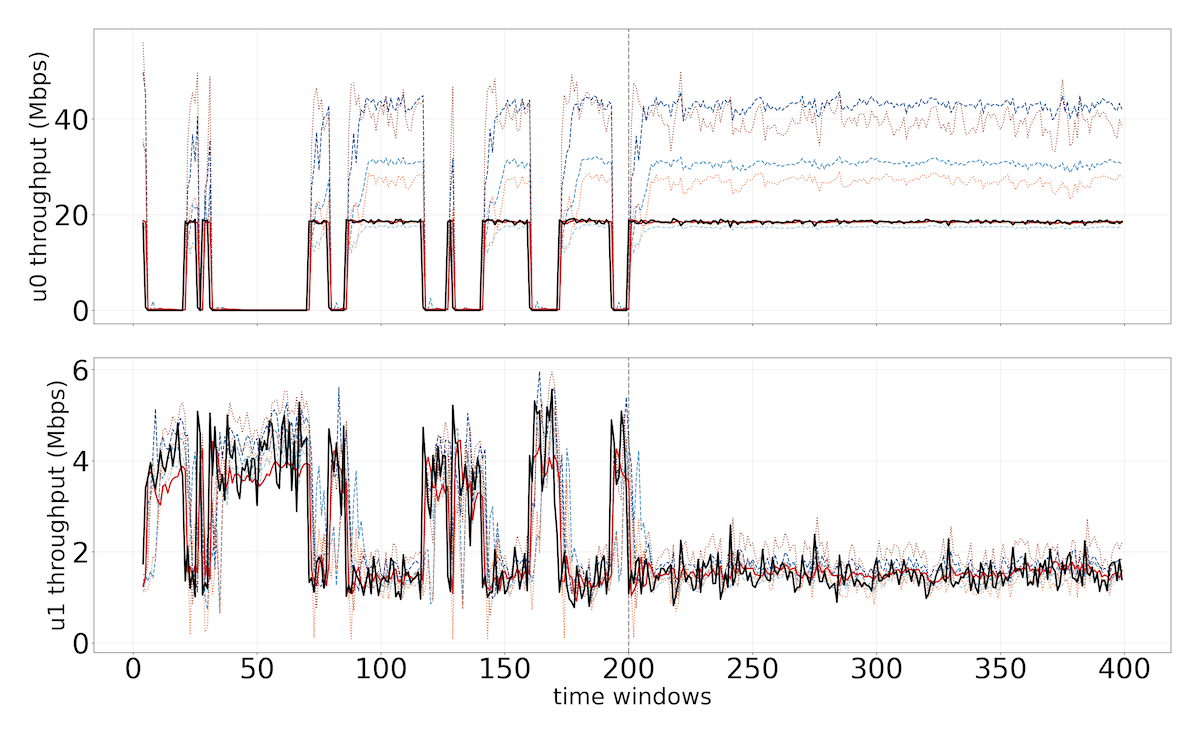}
        \caption{\textit{patternDemand}, $C=20$ Mbps}
    \end{subfigure}
    \hfill
    \begin{subfigure}[t]{0.24\textwidth}
        \centering
        \includegraphics[width=\linewidth]{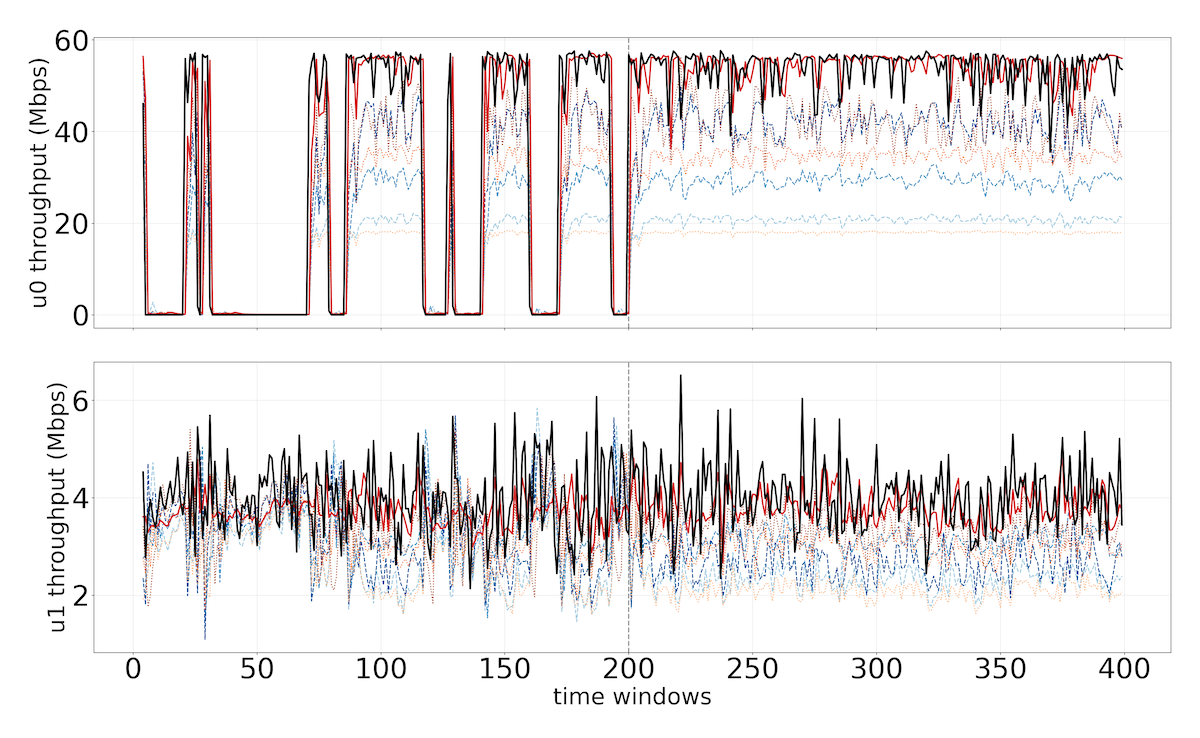}
        \caption{\textit{patternDemand}, $C=60$ Mbps}
    \end{subfigure}
    \hfill
    \begin{subfigure}[t]{0.24\textwidth}
        \centering
        \includegraphics[width=\linewidth]{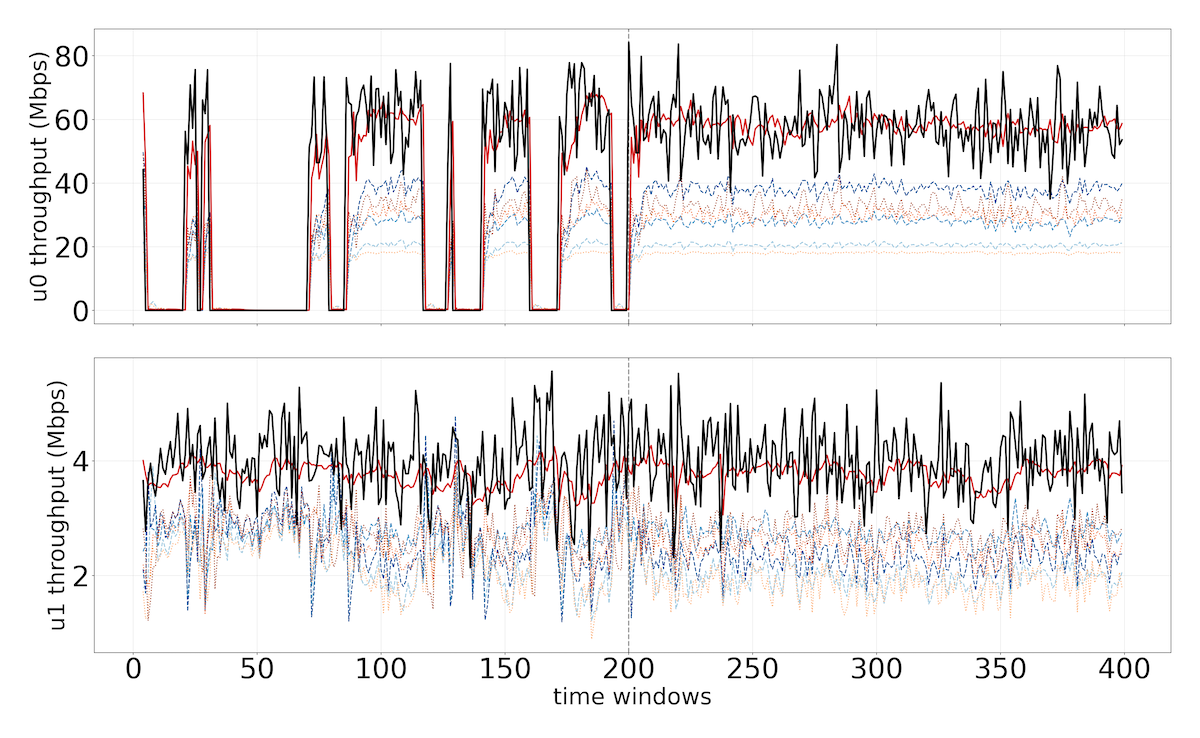}
        \caption{\textit{patternDemand}, $C=280$ Mbps}
    \end{subfigure}
    \hfill
    \begin{subfigure}[t]{0.24\textwidth}
        \centering
        \includegraphics[width=\linewidth]{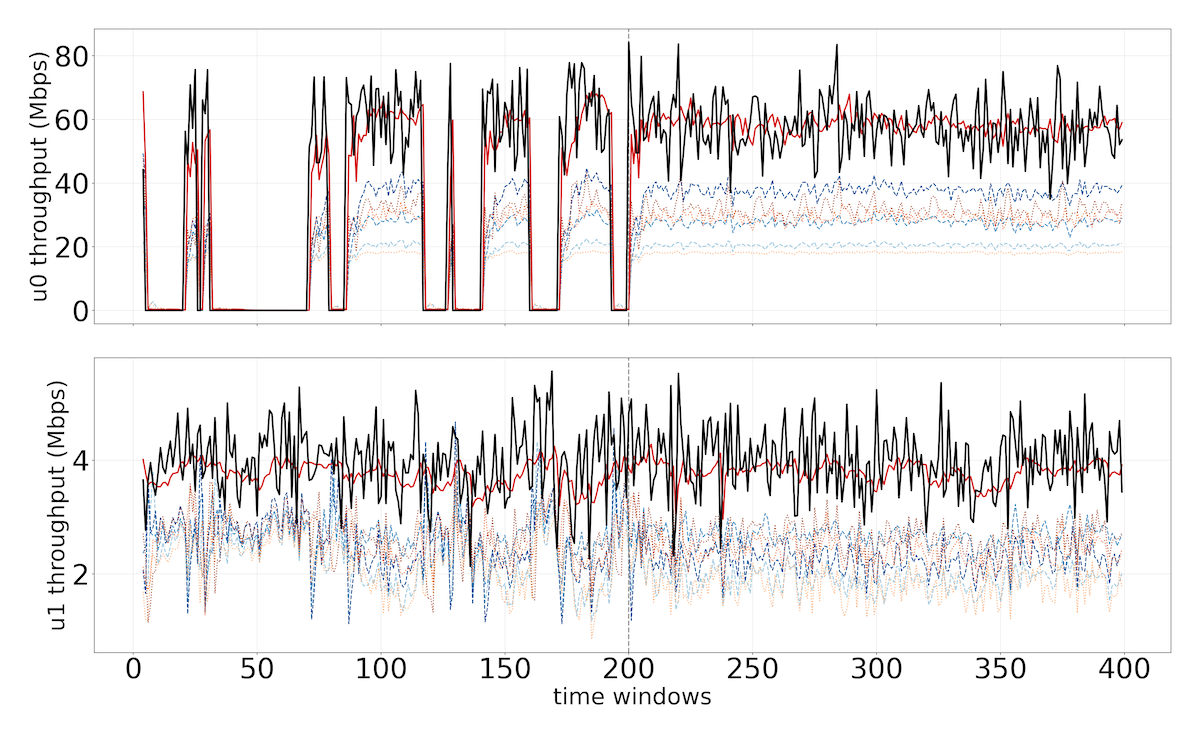}
        \caption{\textit{patternDemand}, $C=360$ Mbps}
    \end{subfigure}

    \vspace{0.3em}
    \includegraphics[width=0.9\textwidth]{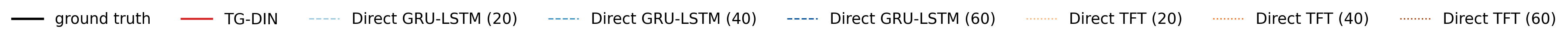}
    \caption{Synthetic throughput traces when the traffic of $u_0$ changes at window 200, marked by the dashed vertical line, under four test capacities.}
    \label{fig:synthetic-timeseries-all}
\end{figure*}

\begin{table}[ht!]
\centering
\caption{Cross-capacity results on the synthetic sweep. RMSE is in Mbps and NMSE is in dB, and lower is better for both, while a higher Pearson $r$ is better. Each metric is computed for each flow and averaged over the two flows. GRU-$c$ and TFT-$c$ denote the GRU-LSTM and TFT-style baselines trained at capacity $c$ Mbps. Bold marks the best model in a column of each block, and a dagger marks a baseline evaluated at the single capacity it was trained on.}
\label{tab:cross-cap-rmse}
\footnotesize
\setlength{\tabcolsep}{2.2pt}
\begin{tabular}{@{}lccccccc@{}}
\toprule
 & \multicolumn{7}{c}{\textbf{Test capacity (Mbps)}} \\
 \cmidrule(lr){2-8}
\textbf{Method} & 20 & 40 & 60 & 120 & 200 & 280 & 360 \\
\midrule
\multicolumn{8}{@{}l}{\textit{RMSE (Mbps)}} \\
GRU-20 & 2.35$^{\dagger}$ & 6.98 & 12.32 & 35.18 & 72.52 & 72.90 & 80.37 \\
GRU-40 & 5.80 & 2.60$^{\dagger}$ & 7.77 & 29.55 & 66.26 & 68.27 & 76.54 \\
GRU-60 & 10.53 & 7.39 & 4.36$^{\dagger}$ & 22.80 & 58.13 & 62.31 & 71.45 \\
TFT-20 & \textbf{2.30}$^{\dagger}$ & 8.21 & 13.77 & 37.75 & 76.06 & 75.59 & 82.74 \\
TFT-40 & 6.00 & 2.52$^{\dagger}$ & 7.89 & 30.63 & 67.87 & 69.36 & 77.79 \\
TFT-60 & 12.03 & 8.05 & 4.24$^{\dagger}$ & 23.65 & 59.84 & 63.49 & 72.88 \\
\cmidrule(lr){1-8}
\textbf{TG-DIN} & 2.32 & \textbf{2.43} & \textbf{4.08} & \textbf{8.46} & \textbf{15.14} & \textbf{15.66} & \textbf{13.65} \\
\midrule
\multicolumn{8}{@{}l}{\textit{NMSE (dB)}} \\
GRU-20 & \textminus{}11.34$^{\dagger}$ & \textminus{}7.25 & \textminus{}4.31 & \textminus{}2.09 & \textminus{}1.25 & \textminus{}0.99 & \textminus{}0.73 \\
GRU-40 & \textminus{}3.59 & \textminus{}16.23$^{\dagger}$ & \textminus{}8.30 & \textminus{}3.60 & \textminus{}2.03 & \textminus{}1.56 & \textminus{}1.15 \\
GRU-60 & 1.66 & \textminus{}6.75 & \textminus{}13.33$^{\dagger}$ & \textminus{}5.86 & \textminus{}3.17 & \textminus{}2.35 & \textminus{}1.75 \\
TFT-20 & \textbf{\textminus{}11.55}$^{\dagger}$ & \textminus{}5.88 & \textminus{}3.35 & \textminus{}1.47 & \textminus{}0.84 & \textminus{}0.68 & \textminus{}0.48 \\
TFT-40 & \textminus{}3.33 & \textminus{}16.43$^{\dagger}$ & \textminus{}8.17 & \textminus{}3.29 & \textminus{}1.83 & \textminus{}1.42 & \textminus{}1.01 \\
TFT-60 & 2.78 & \textminus{}6.05 & \textminus{}13.56$^{\dagger}$ & \textminus{}5.54 & \textminus{}2.92 & \textminus{}2.18 & \textminus{}1.58 \\
\cmidrule(lr){1-8}
\textbf{TG-DIN} & \textminus{}11.48 & \textbf{\textminus{}16.79} & \textbf{\textminus{}13.92} & \textbf{\textminus{}14.48} & \textbf{\textminus{}15.17} & \textbf{\textminus{}14.79} & \textbf{\textminus{}16.50} \\
\midrule
\multicolumn{8}{@{}l}{\textit{Pearson $r$}} \\
GRU-20 & 0.933$^{\dagger}$ & 0.957 & 0.949 & 0.931 & 0.936 & 0.858 & 0.885 \\
GRU-40 & 0.901 & 0.972$^{\dagger}$ & 0.966 & 0.958 & 0.955 & 0.899 & 0.934 \\
GRU-60 & 0.885 & 0.966 & 0.970$^{\dagger}$ & 0.968 & 0.963 & 0.938 & 0.959 \\
TFT-20 & \textbf{0.936}$^{\dagger}$ & 0.941 & 0.931 & 0.898 & 0.896 & 0.790 & 0.814 \\
TFT-40 & 0.895 & 0.974$^{\dagger}$ & 0.968 & 0.957 & 0.950 & 0.891 & 0.930 \\
TFT-60 & 0.863 & 0.966 & 0.971$^{\dagger}$ & 0.969 & 0.964 & 0.931 & 0.955 \\
\cmidrule(lr){1-8}
\textbf{TG-DIN} & 0.935 & \textbf{0.975} & \textbf{0.974} & \textbf{0.976} & \textbf{0.975} & \textbf{0.978} & \textbf{0.985} \\
\bottomrule
\end{tabular}
\end{table}

\subsection{Cross-Capacity Generalization without Adaptation}\label{subsec:cross-capacity-no-adap}

We first compare TG-DIN with the six single-capacity direct baselines on the synthetic test grid. Each baseline is trained at one source capacity (20, 40, or 60~Mbps) and evaluated at all test capacities without target-side adaptation. Fig.~\ref{fig:cross-capacity-main} summarizes RMSE, NMSE, and Pearson $r$ across capacities, and Table~\ref{tab:cross-cap-rmse} gives the values.
The direct baselines are strongly capacity-specific: each performs best near its training capacity and degrades rapidly under capacity shift. The GRU-LSTM baseline trained at 20~Mbps, for example, has an RMSE of 2.35~Mbps at its source capacity and 80.37~Mbps at 360~Mbps, and the corresponding TFT-style baseline rises from 2.30 to 82.74~Mbps. TG-DIN, trained without any collected data at the test capacities, remains stable across the whole range. At the baselines' training capacities, its RMSE matches the corresponding baseline at 20~Mbps (2.32 versus 2.30~Mbps) and is lower at 40 and 60~Mbps (2.43 and 4.08~Mbps). At higher capacities, its RMSE grows with the throughput scale but stays far below the baselines, reaching 13.65~Mbps at 360~Mbps compared with 71.45 to 82.74~Mbps.
NMSE shows that these gains are not merely an effect of throughput scale. We compute it as the sum of squared errors divided by the sum of squared observed throughput, in dB, so 0~dB equals the error of predicting zero throughput and each 10~dB decrease is a tenfold reduction in relative error. TG-DIN stays between $-11.48$ and $-16.79$~dB across the range and has the lowest NMSE at every capacity except 20~Mbps, where it reaches $-11.48$~dB against $-11.55$~dB for the best baseline. Each baseline reaches a comparable value only at its own training capacity, for example $-16.23$~dB for the GRU-LSTM baseline trained at 40~Mbps, and rises toward 0~dB elsewhere, reaching $-0.48$ to $-1.75$~dB at 360~Mbps. At 20~Mbps, both baselines trained at 60~Mbps exceed 0~dB, meaning their squared error is larger than the squared observed throughput. Pearson $r$ is highest for TG-DIN from 40 to 360~Mbps, so it also preserves the temporal structure of per-flow throughput better under capacity shift.

Fig.~\ref{fig:synthetic-timeseries-all} examines whether the models keep tracking throughput when the traffic of $u_0$ changes mid-trace. Before the change, TG-DIN follows the on-off periods of $u_0$ at every capacity, whereas the baselines follow them only near their own training capacity. After the change, all models need a few windows to react, and their error settles at a new level after about 20 windows, with TG-DIN settling closer to the ground truth. At 280~Mbps, its RMSE over these settled windows is 16.60~Mbps against 23.02~Mbps for the best baseline in \textit{demandOnly}, and 3.31 against 6.52~Mbps in \textit{patternDemand}. Over the whole trace at 280~Mbps, TG-DIN reaches 13.56, 6.21, and 6.54~Mbps in \textit{demandOnly}, \textit{patternOnly}, and \textit{patternDemand}, compared with 18.28, 7.57, and 8.76~Mbps for the best baseline. At 20~Mbps the link is the bottleneck and throughput is pinned near capacity, so all models are close. Even there, the baselines trained at 40 and 60~Mbps predict throughput far above the link capacity, whereas the theory layer keeps TG-DIN below it.

Because the same demand traces are replayed at all four capacities, the observed throughput differs across them only because capacity differs, which reveals the source of this gap. In \textit{demandOnly}, the mean observed throughput of $u_0$ rises from 5.54~Mbps at 20~Mbps to 25.66~Mbps at 280 and 360~Mbps, and TG-DIN follows it, from 5.40 to 23.34~Mbps. Each baseline instead predicts nearly the same level at all four capacities, set by its training capacity: about 5~Mbps for the baselines trained at 20~Mbps, 8 to 10~Mbps for those trained at 40~Mbps, and 11 to 14~Mbps for those trained at 60~Mbps. Across the four capacities, a baseline's prediction moves by at most 3.1~Mbps, while the observed throughput moves by 20.1~Mbps. In \textit{patternOnly} and \textit{patternDemand}, some baselines even move in the wrong direction; the GRU-LSTM baseline trained at 60~Mbps, for example, drops from 16.18 to 10.85~Mbps while the observed throughput rises from 9.76 to 16.15~Mbps.


\begin{figure*}[ht!]
    \centering
    \begin{subfigure}[t]{0.32\textwidth}
        \centering
        \includegraphics[width=\linewidth]{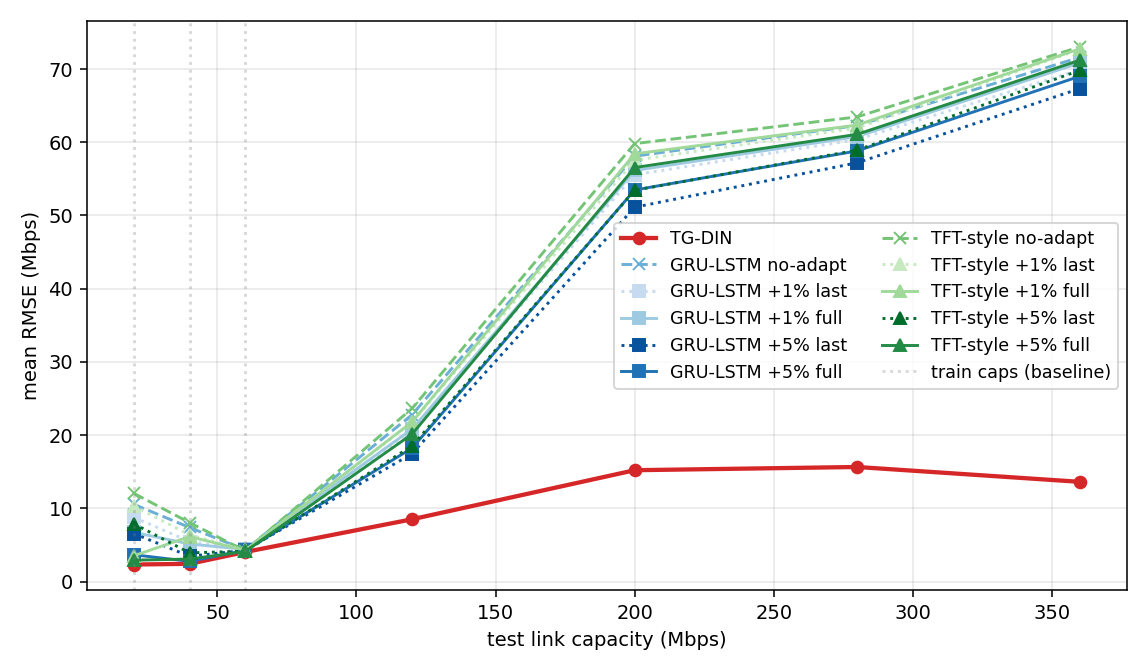}
        \caption{RMSE.}
    \end{subfigure}
    \hfill
    \begin{subfigure}[t]{0.32\textwidth}
        \centering
        \includegraphics[width=\linewidth]{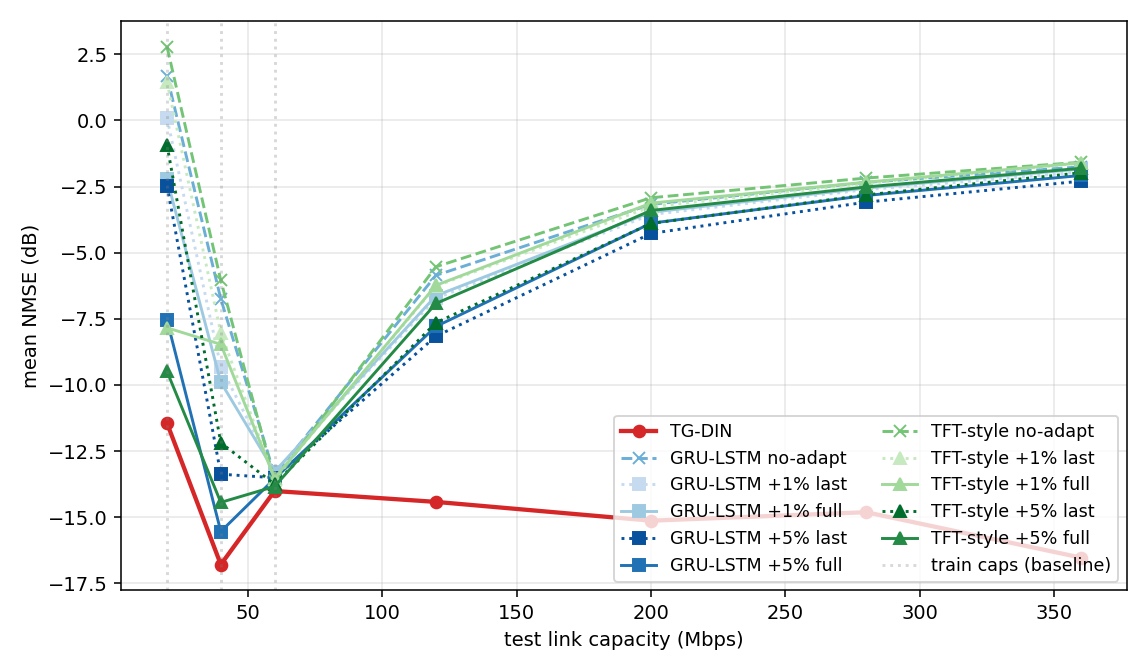}
        \caption{NMSE (dB).}
    \end{subfigure}
    \hfill
    \begin{subfigure}[t]{0.32\textwidth}
        \centering
        \includegraphics[width=\linewidth]{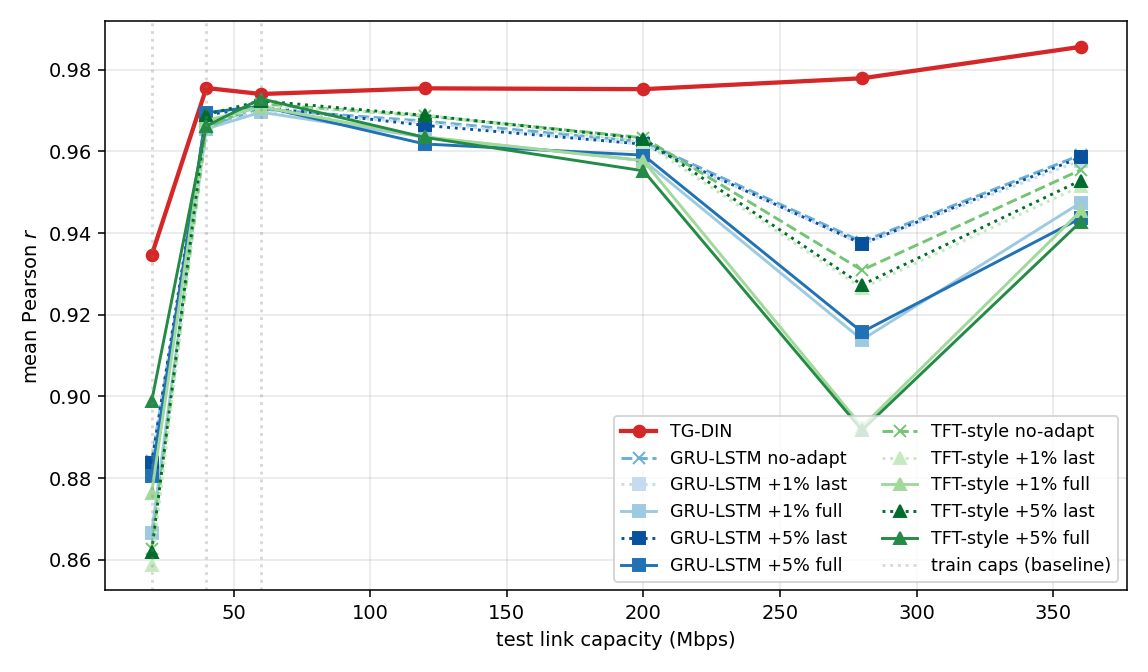}
        \caption{Pearson $r$.}
    \end{subfigure}
    \caption{Target-adapted fine-tuning under capacity drift. Direct baselines are adapted using 1\% or 5\% target-capacity calibration data with either full-model or last-layer updates.}
    \label{fig:finetune-main}
\end{figure*}

\begin{table*}[ht!]
\centering
\caption{Fine-tuning under capacity shift. GRU-60 and TFT-60 are adapted to each test capacity (Mbps) using 1\% or 5\% of its windows, updating the output layer (last) or all parameters (full); TG-DIN is not adapted. All models are evaluated on the remaining windows. Best results in bold.}
\label{tab:ft-rmse}
\footnotesize
\setlength{\tabcolsep}{2.3pt}
\begin{tabular}{@{}ll *{6}{c} @{\hspace{7pt}} *{6}{c} @{\hspace{7pt}} *{6}{c}@{}}
\toprule
 & & \multicolumn{6}{c}{\textbf{RMSE (Mbps)}} & \multicolumn{6}{c}{\textbf{NMSE (dB)}} & \multicolumn{6}{c}{\textbf{Pearson $r$}} \\
\cmidrule(lr){3-8}\cmidrule(lr){9-14}\cmidrule(l){15-20}
\textbf{Model} & \textbf{Method} & 20 & 40 & 120 & 200 & 280 & 360 & 20 & 40 & 120 & 200 & 280 & 360 & 20 & 40 & 120 & 200 & 280 & 360 \\
\midrule
\multirow{5}{*}{GRU-60} & No-adapt & 10.54 & 7.39 & 22.80 & 58.10 & 62.29 & 71.60 & 1.66 & \textminus6.75 & \textminus5.86 & \textminus3.17 & \textminus2.35 & \textminus1.75 & 0.884 & 0.966 & 0.967 & 0.962 & 0.938 & 0.959 \\
 & +1\% last & 8.77 & 5.47 & 20.63 & 55.56 & 60.37 & 69.90 & 0.08 & \textminus9.31 & \textminus6.72 & \textminus3.56 & \textminus2.62 & \textminus1.95 & 0.883 & 0.967 & 0.967 & 0.962 & 0.937 & 0.958 \\
 & +1\% full & 6.73 & 5.14 & 20.85 & 56.15 & 60.78 & 70.84 & \textminus2.23 & \textminus9.90 & \textminus6.63 & \textminus3.47 & \textminus2.56 & \textminus1.84 & 0.867 & 0.966 & 0.963 & 0.958 & 0.914 & 0.947 \\
 & +5\% last & 6.55 & 3.46 & 17.47 & 51.13 & 57.20 & 67.33 & \textminus2.47 & \textminus13.37 & \textminus8.16 & \textminus4.27 & \textminus3.09 & \textminus2.30 & 0.884 & 0.969 & 0.966 & 0.962 & 0.937 & 0.959 \\
 & +5\% full & 3.66 & 2.77 & 18.23 & 53.48 & 58.82 & 69.01 & \textminus7.55 & \textminus15.54 & \textminus7.79 & \textminus3.89 & \textminus2.84 & \textminus2.08 & 0.881 & 0.970 & 0.962 & 0.959 & 0.916 & 0.944 \\
\midrule
\multirow{5}{*}{TFT-60} & No-adapt & 12.04 & 8.05 & 23.65 & 59.82 & 63.47 & 73.03 & 2.78 & \textminus6.05 & \textminus5.53 & \textminus2.92 & \textminus2.18 & \textminus1.58 & 0.863 & 0.966 & 0.969 & 0.963 & 0.931 & 0.955 \\
 & +1\% last & 10.32 & 6.38 & 21.74 & 57.55 & 61.94 & 72.68 & 1.44 & \textminus8.03 & \textminus6.26 & \textminus3.26 & \textminus2.39 & \textminus1.61 & 0.859 & 0.967 & 0.969 & 0.962 & 0.927 & 0.951 \\
 & +1\% full & 3.53 & 6.15 & 21.81 & 58.43 & 62.35 & 72.76 & \textminus7.84 & \textminus8.47 & \textminus6.23 & \textminus3.13 & \textminus2.34 & \textminus1.61 & 0.876 & 0.968 & 0.964 & 0.958 & 0.893 & 0.945 \\
 & +5\% last & 7.89 & 3.93 & 18.53 & 53.44 & 58.93 & 69.82 & \textminus0.93 & \textminus12.19 & \textminus7.65 & \textminus3.89 & \textminus2.81 & \textminus1.98 & 0.862 & 0.969 & 0.969 & 0.963 & 0.927 & 0.953 \\
 & +5\% full & 2.92 & 3.08 & 20.15 & 56.54 & 61.08 & 71.20 & \textminus9.49 & \textminus14.44 & \textminus6.92 & \textminus3.40 & \textminus2.52 & \textminus1.81 & 0.899 & 0.966 & 0.963 & 0.955 & 0.892 & 0.943 \\
\midrule
\textbf{TG-DIN} & No-adapt & \textbf{2.33} & \textbf{2.43} & \textbf{8.51} & \textbf{15.21} & \textbf{15.65} & \textbf{13.62} & \textbf{\textminus11.45} & \textbf{\textminus16.79} & \textbf{\textminus14.42} & \textbf{\textminus15.13} & \textbf{\textminus14.81} & \textbf{\textminus16.53} & \textbf{0.935} & \textbf{0.976} & \textbf{0.975} & \textbf{0.975} & \textbf{0.978} & \textbf{0.986} \\
\bottomrule
\end{tabular}
\end{table*}

\subsection{Comparison with Target-Adapted Fine-Tuning}\label{subsec:adap}

We next examine whether limited target-capacity data can close the generalization gap of the direct baselines. Both baselines are initialized from their 60~Mbps checkpoints and fine-tuned for each target capacity in $\{20,40,120,200,280,360\}$~Mbps. The calibration set is a random 1\% or 5\% of the test windows at that capacity, and all models, including TG-DIN, are evaluated on the remaining windows. We consider two strategies: \emph{full} fine-tuning of all parameters and \emph{last-layer} fine-tuning of the output head only. As demonstrated in Fig.~\ref{fig:finetune-main} and Table~\ref{tab:ft-rmse}, fine-tuning helps most at low capacities for the baselines. With 5\% full fine-tuning at 20~Mbps, the RMSE of the GRU-LSTM baseline falls from 10.54 to 3.66~Mbps and that of the TFT-style baseline from 12.04 to 2.92~Mbps. At 20 and 40~Mbps, full fine-tuning outperforms last-layer fine-tuning for both baselines and both budgets. Even so, the best adapted baselines at these capacities (2.92 and 2.77~Mbps) remain above TG-DIN (2.33 and 2.43~Mbps), which uses no target data.
At higher capacities, however, fine-tuning recovers only a small part of the gap. The best adapted baseline, the GRU-LSTM baseline with 5\% last-layer fine-tuning, reaches 17.47, 51.13, 57.20, and 67.33~Mbps at 120, 200, 280, and 360~Mbps, compared with 8.51, 15.21, 15.65, and 13.62~Mbps for TG-DIN, respectively. In this regime, last-layer fine-tuning is slightly better than full fine-tuning, but neither strategy brings the baselines close to TG-DIN. TG-DIN therefore achieves the lowest RMSE at every target capacity without any fine-tuning. A small amount of target data can correct a direct predictor near the source condition, but it does not substitute for coverage of the operating range, which TG-DIN obtains through its theory layer.

\subsection{Real-Traffic Transfer}\label{subsec:real}

We next apply the same models, without retraining or fine-tuning, to the real captures collected with the testbed. We report three groups of captures: fixed capacities, a change of application at a fixed capacity, and capacity changes within one session. In all captures, flow $u_1$ runs FaceTime.

\begin{figure*}[ht!]
    \centering
    \begin{subfigure}[t]{0.24\linewidth}
        \centering
        \includegraphics[width=\linewidth]{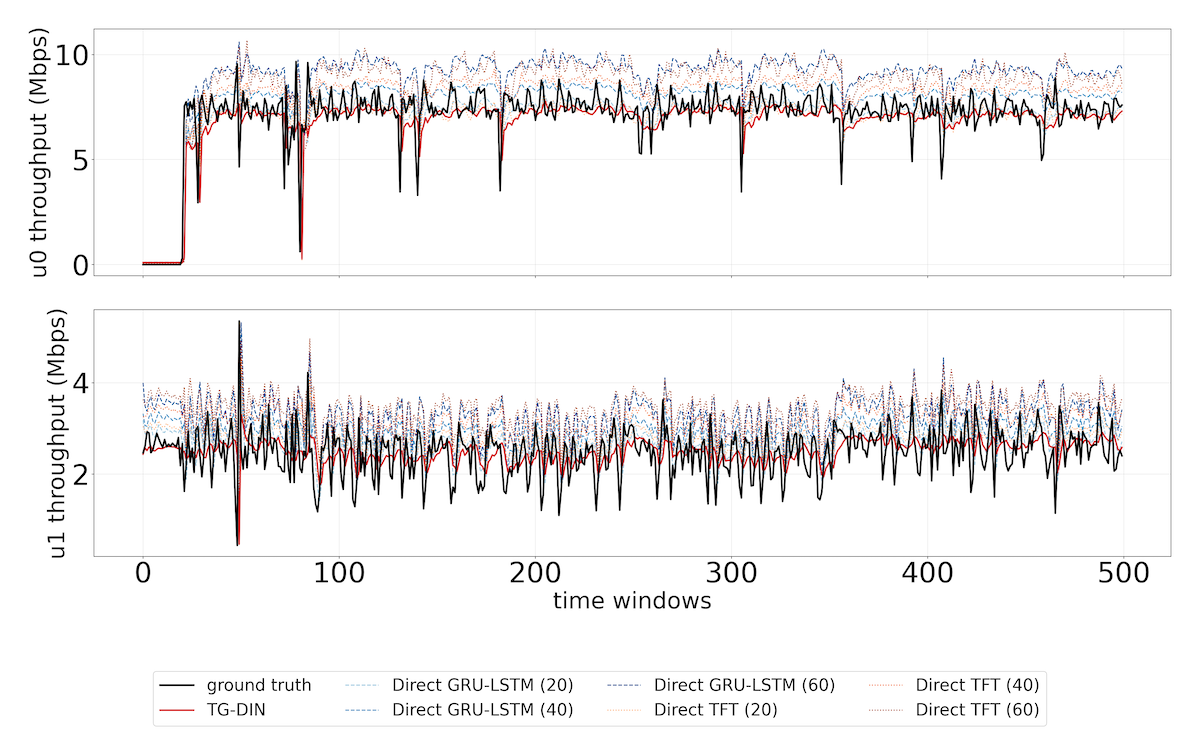}
        \caption{$C = 10$ Mbps}
    \end{subfigure}
    \begin{subfigure}[t]{0.24\linewidth}
        \centering
        \includegraphics[width=\linewidth]{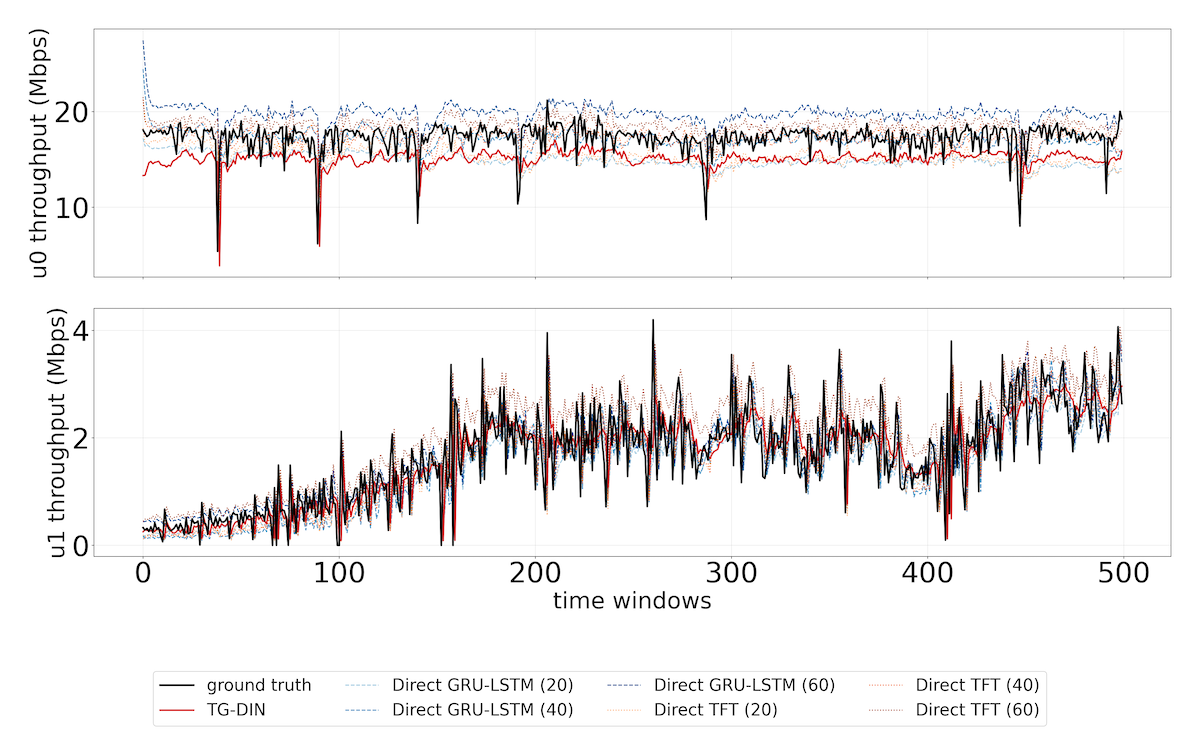}
        \caption{$C = 20$ Mbps}
    \end{subfigure}
    \begin{subfigure}[t]{0.24\linewidth}
        \centering
        \includegraphics[width=\linewidth]{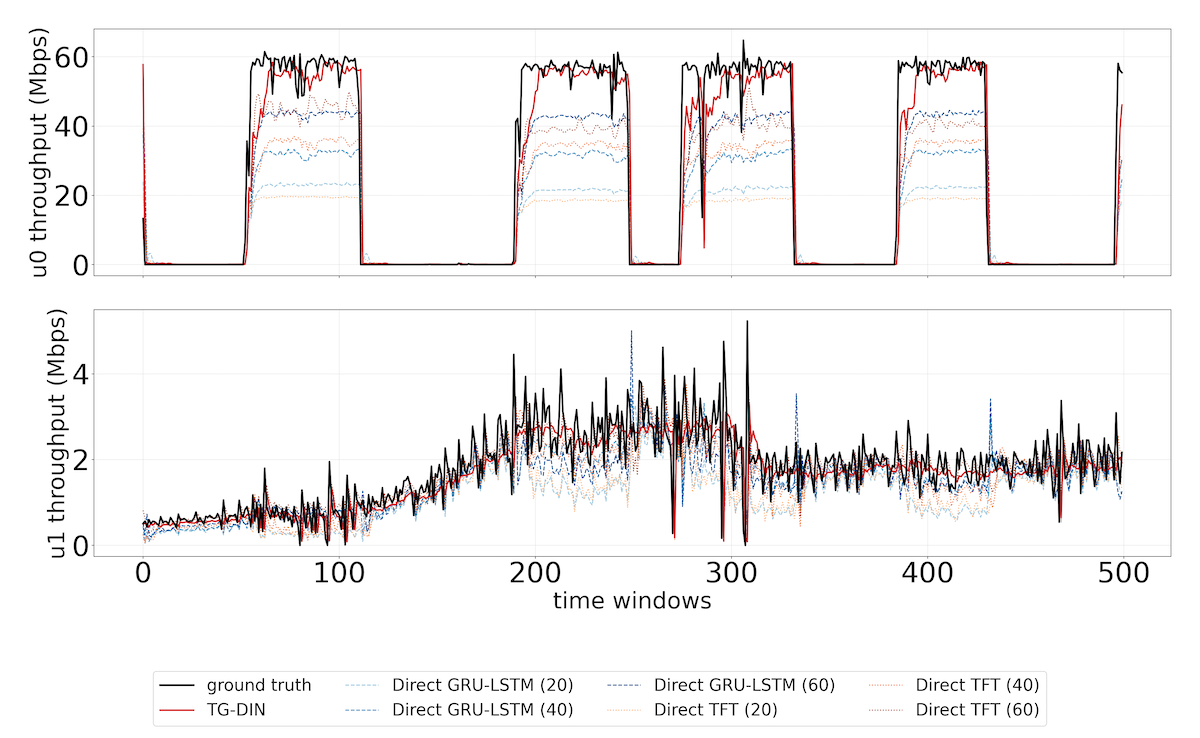}
        \caption{$C = 60$ Mbps}
    \end{subfigure}
    \begin{subfigure}[t]{0.24\linewidth}
        \centering
        \includegraphics[width=\linewidth]{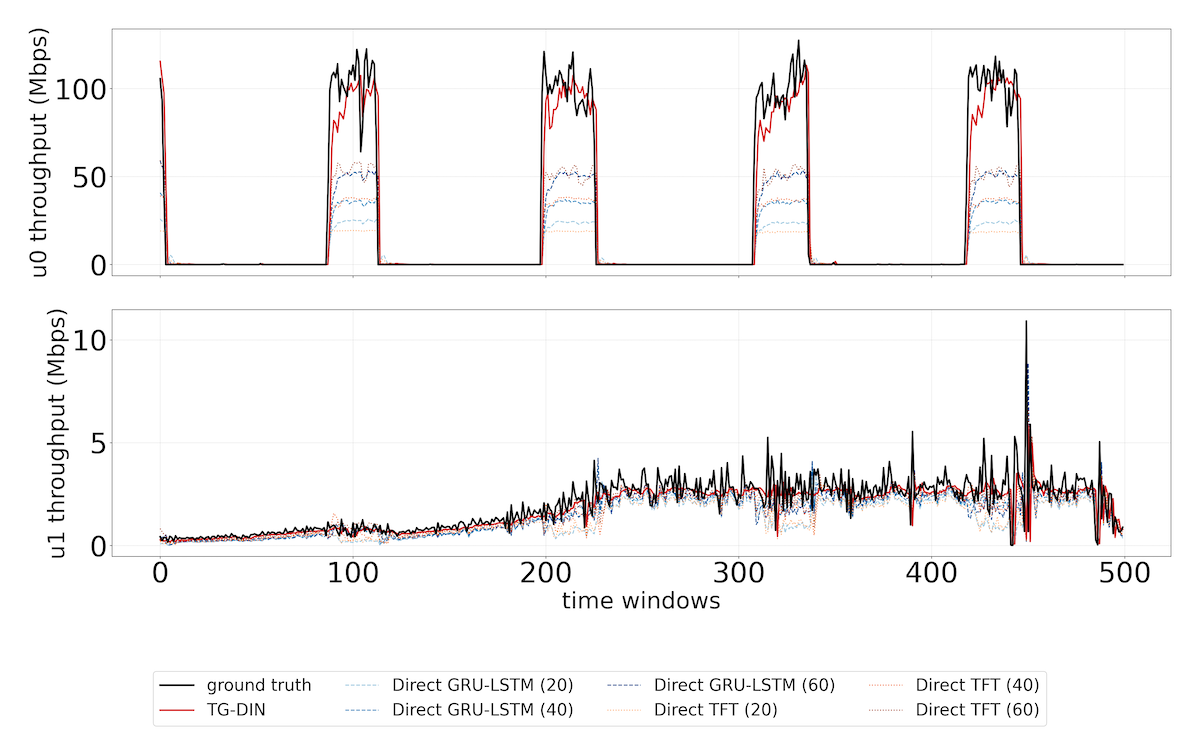}
        \caption{$C = 120$ Mbps}
    \end{subfigure}
    \caption{Real-traffic throughput at fixed bottleneck capacities. Black: observed; red: TG-DIN; other curves: direct baselines.}
    \label{fig:real_timeseries}
\end{figure*}

\begin{table*}[ht!]
\centering
\caption{Real-traffic evaluation. Capacity columns: $u_0$ streams YouTube and $u_1$ runs FaceTime at four bottleneck capacities. App.\ columns: at 10~Mbps, $u_0$ streams YouTube (YT) or downloads an ISO image. GRU-$c$ and TFT-$c$ are baselines trained at $c$~Mbps; $^{\dagger}$ marks evaluation at the training capacity. Best results in bold.}
\label{tab:real-traffic}
\footnotesize
\setlength{\tabcolsep}{2.4pt}
\begin{tabular}{@{}l *{6}{c} @{\hspace{7pt}} *{6}{c} @{\hspace{7pt}} *{6}{c}@{}}
\toprule
 & \multicolumn{6}{c}{\textbf{RMSE (Mbps)}} & \multicolumn{6}{c}{\textbf{MAE (Mbps)}} & \multicolumn{6}{c}{\textbf{NMSE (dB)}} \\
\cmidrule(lr){2-7}\cmidrule(lr){8-13}\cmidrule(l){14-19}
 & \multicolumn{4}{c}{Capacity (Mbps)} & \multicolumn{2}{c}{App.} & \multicolumn{4}{c}{Capacity (Mbps)} & \multicolumn{2}{c}{App.} & \multicolumn{4}{c}{Capacity (Mbps)} & \multicolumn{2}{c}{App.} \\
\cmidrule(lr){2-5}\cmidrule(lr){6-7}\cmidrule(lr){8-11}\cmidrule(lr){12-13}\cmidrule(lr){14-17}\cmidrule(l){18-19}
\textbf{Method} & 10 & 20 & 60 & 120 & YT & ISO & 10 & 20 & 60 & 120 & YT & ISO & 10 & 20 & 60 & 120 & YT & ISO \\
\midrule
GRU-20 & \textbf{0.67} & 1.86$^{\dagger}$ & 12.49 & 19.87 & 1.35 & 1.33 & \textbf{0.43} & 1.42$^{\dagger}$ & 8.57 & 9.79 & 1.03 & 1.15 & \textbf{\textminus15.76} & \textminus10.97$^{\dagger}$ & \textminus5.18 & \textminus3.61 & \textminus11.14 & \textminus11.56 \\
GRU-40 & 0.87 & \textbf{1.38} & 9.49 & 17.39 & 1.75 & 2.03 & 0.63 & 0.86 & 6.43 & 8.53 & 1.45 & 1.87 & \textminus13.53 & \textbf{\textminus12.65} & \textminus7.76 & \textminus4.97 & \textminus8.85 & \textminus7.79 \\
GRU-60 & 1.41 & 1.87 & 6.56$^{\dagger}$ & 14.34 & 2.64 & 3.52 & 1.15 & 1.32 & 4.19$^{\dagger}$ & 7.01 & 2.34 & 3.34 & \textminus9.60 & \textminus10.78 & \textminus9.11$^{\dagger}$ & \textminus5.51 & \textminus5.27 & \textminus2.95 \\
TFT-20 & 0.70 & 1.77$^{\dagger}$ & 13.39 & 20.92 & 1.64 & 1.90 & 0.47 & 1.30$^{\dagger}$ & 9.17 & 10.29 & 1.35 & 1.76 & \textminus15.28 & \textminus11.16$^{\dagger}$ & \textminus5.05 & \textminus3.43 & \textminus9.44 & \textminus8.23 \\
TFT-40 & 1.05 & 1.39 & 8.61 & 16.98 & 1.90 & 2.35 & 0.80 & \textbf{0.84} & 5.80 & 8.35 & 1.60 & 2.20 & \textminus11.88 & \textminus12.56 & \textminus7.95 & \textminus4.98 & \textminus8.18 & \textminus6.44 \\
TFT-60 & 1.39 & 1.56 & 6.69$^{\dagger}$ & 13.49 & 2.16 & 2.45 & 1.14 & 1.01 & 4.31$^{\dagger}$ & 6.59 & 1.87 & 2.29 & \textminus9.43 & \textminus11.58 & \textminus9.57$^{\dagger}$ & \textminus6.25 & \textminus7.06 & \textminus5.96 \\
\midrule
\textbf{TG-DIN} & 0.68 & 1.73 & \textbf{4.63} & \textbf{7.70} & \textbf{1.16} & \textbf{0.71} & 0.44 & 1.28 & \textbf{1.99} & \textbf{2.80} & \textbf{0.80} & \textbf{0.31} & \textminus15.74 & \textminus11.40 & \textbf{\textminus11.33} & \textbf{\textminus8.88} & \textbf{\textminus12.58} & \textbf{\textminus17.41} \\
\bottomrule
\end{tabular}
\vspace{-2mm}
\end{table*}

\textbf{Fixed Capacities:}
We collect four captures of about 5 minutes each, at 10, 20, 60, and 120~Mbps, with $u_0$ streaming YouTube. For each capture, we compare predicted and observed per-flow throughput over the full trace and report RMSE, MAE, and NMSE (Fig.~\ref{fig:real_timeseries}, Table~\ref{tab:real-traffic}). At 10 and 20~Mbps, which lie at or near the baselines' training capacities, TG-DIN is competitive with the baselines: its RMSE is 0.68~Mbps against 0.67~Mbps for the best baseline at 10~Mbps, and 1.73~Mbps against 1.38~Mbps at 20~Mbps. As capacity moves away from the training capacities, the baselines degrade and TG-DIN achieves the lowest error. At 60~Mbps, TG-DIN reaches 4.63~Mbps while the baselines range from 6.56 to 13.39~Mbps, and at 120~Mbps it reaches 7.70~Mbps while they range from 13.49 to 20.92~Mbps. MAE and NMSE show the same pattern. Fig.~\ref{fig:real_timeseries} shows the source of the gap: at 60 and 120~Mbps, the baselines underestimate the large bursts of $u_0$, whereas TG-DIN follows their level. TG-DIN thus transfers from generated to real traces and, unlike the baselines, remains accurate when the real capacity differs from any capacity seen by a single-condition model.

\begin{figure}[!t]
    \centering
    \begin{subfigure}[t]{0.95\linewidth}
        \centering
        \includegraphics[width=\linewidth]{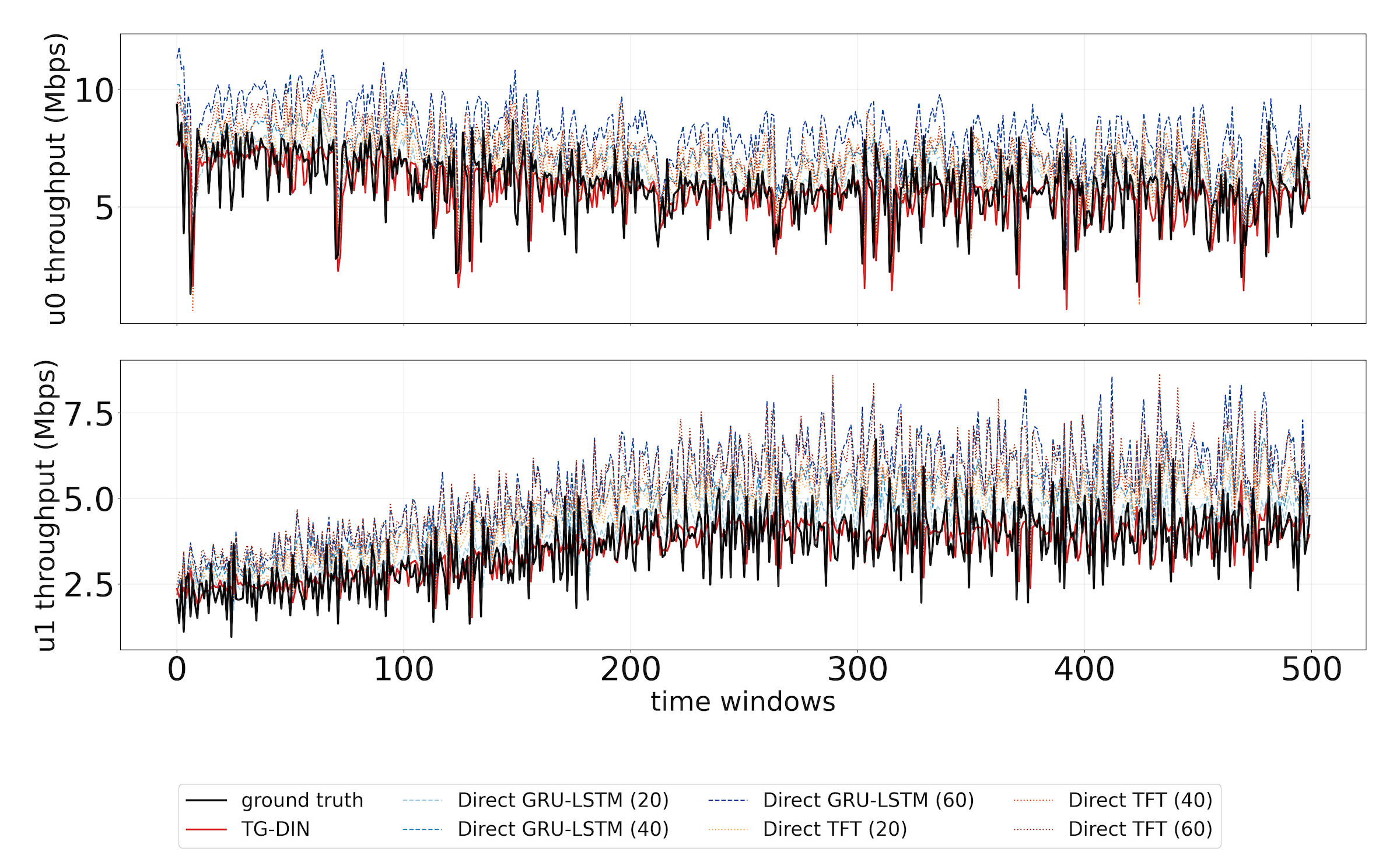}
        \caption{YouTube + FaceTime}
    \end{subfigure}
    \hspace{0.04\textwidth}
    \begin{subfigure}[t]{0.95\linewidth}
        \centering
        \includegraphics[width=\linewidth]{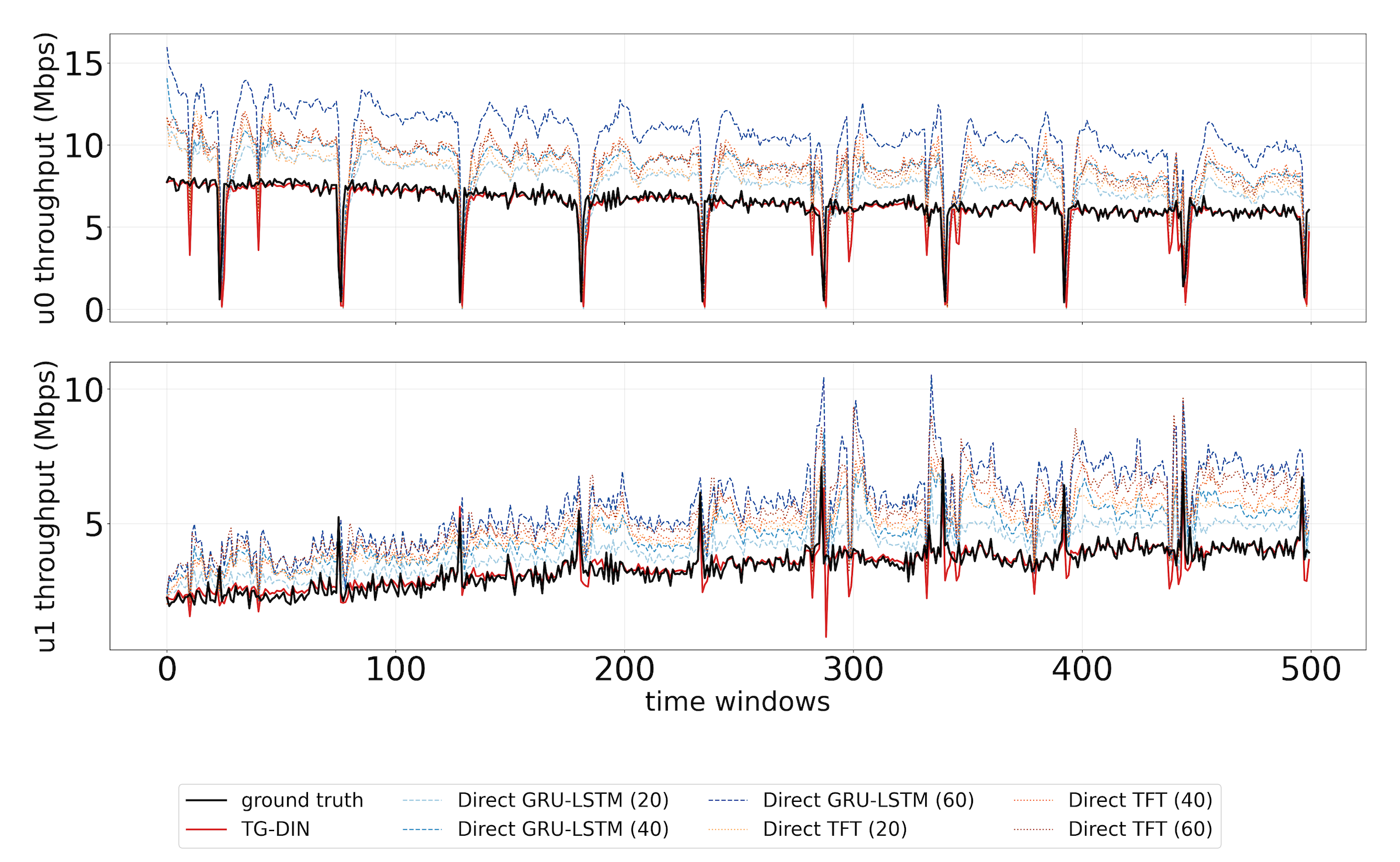}
        \caption{ISO download + FaceTime}
    \end{subfigure}
   \caption{Real-traffic throughput at 10~Mbps with different applications for $u_0$. Black: observed; red: TG-DIN; other curves: direct baselines.}
    \label{fig:real_school}
\end{figure}

\textbf{Application Drift:}
We collect two further captures under a 10~Mbps bottleneck. In the first, $u_0$ streams YouTube, as above. In the second, $u_0$ instead downloads an Ubuntu ISO image over TCP Cubic, which produces sustained traffic rather than the bursty traffic of video streaming. Neither TG-DIN nor the baselines saw TCP traffic during training.
Table~\ref{tab:real-traffic} and Fig.~\ref{fig:real_school} show the results. TG-DIN has the lowest RMSE on both captures, 1.16 and 0.71~Mbps, which is 14\% and 47\% lower than the best baseline (1.35 and 1.33~Mbps), and it also has the lowest MAE and NMSE on both. Because 10~Mbps lies below the baselines' training capacities, part of their error reflects capacity mismatch, so the more informative comparison is how each model's error changes between the two applications at the same capacity. When $u_0$ switches from YouTube to the ISO download, the error of five of the six baselines increases, while the error of TG-DIN decreases from 1.16 to 0.71~Mbps. In Fig.~\ref{fig:real_school}, most baselines overestimate the download flow, whereas TG-DIN follows its level. This is consistent with the theory layer being agnostic to how a flow's offered rate is produced.

\begin{figure}[!t]
    \centering
    \includegraphics[width=0.99\linewidth]{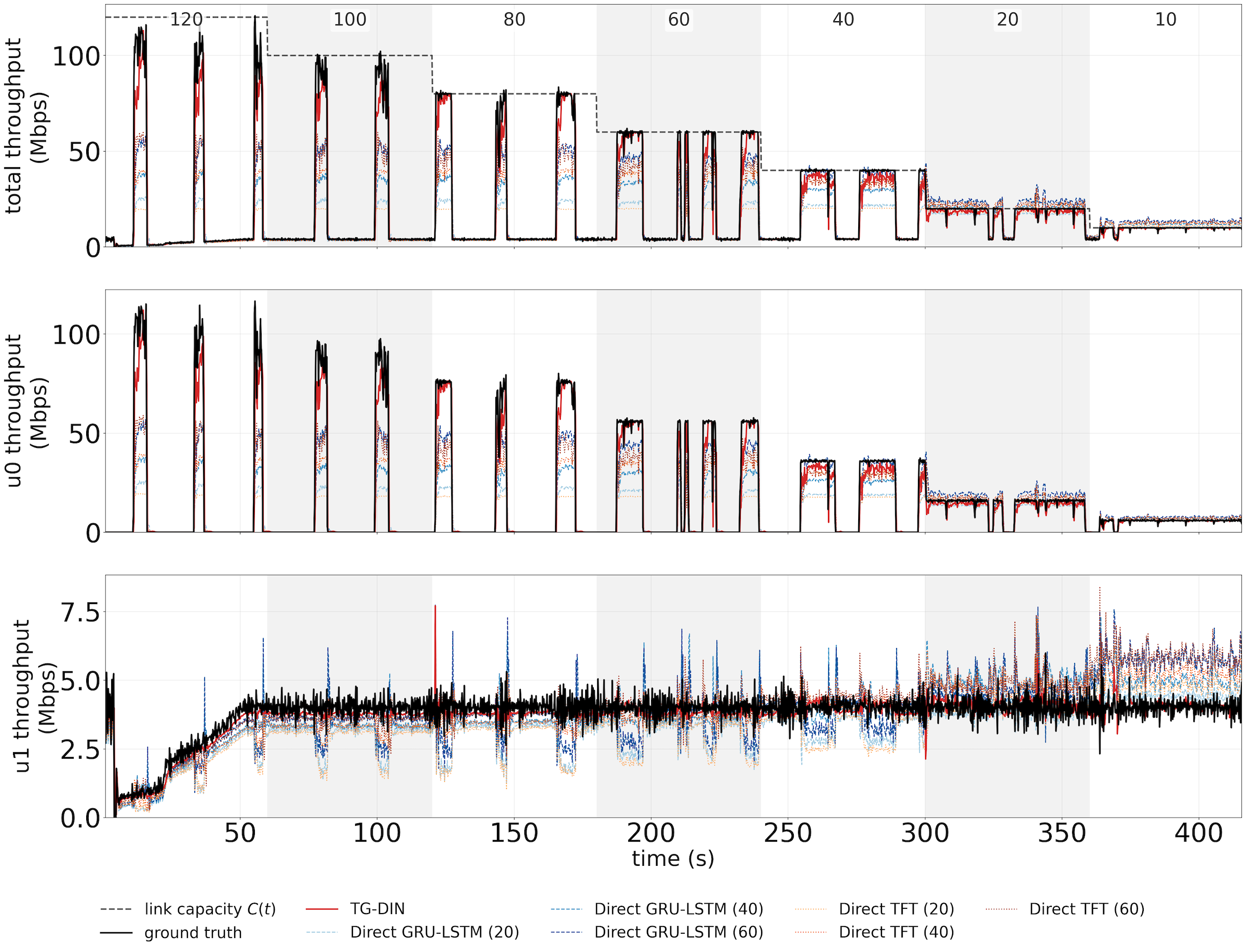}
    \caption{Real-traffic throughput while the bottleneck capacity is stepped down from 120 to 10 Mbps in seven stages of one minute each.}
    \label{fig:ladder-real}
\end{figure}

\textbf{Dynamic Capacity Changes:}
In one session, the bottleneck capacity steps down from 120~Mbps through 100, 80, 60, 40, and 20 to 10~Mbps, with each stage lasting about one minute, while $u_0$ streams YouTube at 4K resolution. As the capacity drops, the bursts of $u_0$ are cut to the new capacity at each stage, and TG-DIN follows the new plateau within each stage (Fig.~\ref{fig:ladder-real}). The baselines underestimate these bursts from 120 to 60~Mbps, capacities at or above their training range, and most of them overestimate the smaller flow $u_1$ at 10~Mbps.
Table~\ref{tab:ladder-real} lists the RMSE within each stage. TG-DIN has the lowest error in six of the seven stages and over the whole run, 6.00~Mbps against 9.77 to 16.03~Mbps for the baselines. The exception is the 40~Mbps stage, where the GRU-LSTM baseline trained at 60~Mbps is slightly lower (3.43 versus 3.73~Mbps). The advantage is largest at high capacity, for example 9.98~Mbps against 16.29 to 25.08~Mbps at 120~Mbps. Over the whole run, MAE and NMSE give the same ranking, with 1.59~Mbps and $-11.22$~dB for TG-DIN compared with 3.26~Mbps and $-6.99$~dB for the best baseline. This session reflects a practical case in which the network condition changes during operation: a direct predictor would require traces from each capacity level to follow such changes, whereas TG-DIN only needs the current capacity as an input.

\begin{table}[!t]
\centering
\caption{Real-traffic capacity ladder: RMSE (Mbps) within each capacity stage, and RMSE, MAE (Mbps), and NMSE (dB) over the whole run. Best results in bold.}
\label{tab:ladder-real}
\footnotesize
\setlength{\tabcolsep}{1.8pt}
\begin{tabular}{@{}lcccccccccc@{}}
\toprule
 & \multicolumn{7}{c}{\textbf{RMSE per capacity stage (Mbps)}} & \multicolumn{3}{c}{\textbf{Whole run}} \\
\cmidrule(lr){2-8}\cmidrule(lr){9-11}
\textbf{Method} & 120 & 100 & 80 & 60 & 40 & 20 & 10 & RMSE & MAE & NMSE \\
\midrule
GRU-20 & 23.91 & 18.12 & 19.48 & 15.70 & 8.67 & 2.23 & 0.76 & 15.21 & 5.14 & \textminus{}3.15 \\
GRU-40 & 21.02 & 15.68 & 16.32 & 12.49 & 5.92 & 1.98 & 1.23 & 12.87 & 4.20 & \textminus{}4.60 \\
GRU-60 & 17.08 & 12.00 & 11.54 & 8.20 & \textbf{3.43} & 3.45 & 1.93 & 9.77 & 3.26 & \textminus{}6.99 \\
TFT-20 & 25.08 & 19.05 & 20.75 & 16.62 & 9.09 & 1.94 & 0.90 & 16.03 & 5.39 & \textminus{}2.69 \\
TFT-40 & 20.15 & 14.55 & 14.71 & 10.35 & 4.24 & 2.31 & 1.43 & 11.79 & 3.80 & \textminus{}5.36 \\
TFT-60 & 16.29 & 12.05 & 12.34 & 9.42 & 4.70 & 2.63 & 1.67 & 9.92 & 3.35 & \textminus{}6.86 \\
\midrule
\textbf{TG-DIN} & \textbf{9.98} & \textbf{7.36} & \textbf{6.32} & \textbf{6.24} & 3.73 & \textbf{1.91} & \textbf{0.61} & \textbf{6.00} & \textbf{1.59} & \textbf{\textminus{}11.22} \\
\bottomrule
\end{tabular}
\end{table}

\subsection{Complexity and Cost Analysis}

Table~\ref{tab:complexity} compares the parameters, the multiply-accumulate operations (MACs) needed to process one window with two flows, and the traffic that must be collected to train each model.
The demand predictor of TG-DIN has 102,529 parameters (0.40~MiB), between the TFT-style baseline (75,074) and the GRU-LSTM baseline (132,290); the theory layer adds no trainable parameters and only a few elementwise operations per flow. For one window with two flows, TG-DIN needs about 171K MACs, against about 355K for the TFT-style baseline and 621K for the GRU-LSTM baseline, which run recurrent layers over all $K$ steps. Because the encoder and head are applied per flow and the shared context is computed once, the cost of TG-DIN grows linearly with the number of flows, reaching about 448K MACs for six flows. All three models stay below one million MACs per window, so the difference in compute is small in absolute terms.

The main difference lies in obtaining training data. Each direct baseline is trained on 120 traces at one capacity, about 2.7 hours of traffic that would have to be collected in deployment, and it performs well only near that capacity (Table~\ref{tab:cross-cap-rmse}). Covering the seven capacities of our synthetic sweep in this way would require about 19~hours of collected traffic, and each new capacity, traffic pattern, or application would require another round of collection and retraining. Fine-tuning reduces the amount of new data but still requires traces from the target condition, and it does not close the gap at high capacities. TG-DIN requires no collected traces, because its training data are generated by running the theory layer forward. Its complete training run used 53 rounds (10,600 generated traces), stopped early after 10 rounds without improvement, and finished in about 36 minutes on a single GPU. If a deployment falls outside the generated range, TG-DIN may also lose accuracy; the generation range can then be extended and the model retrained, still without collecting new data.

\begin{table}[!t]
\centering
\caption{Model complexity and training-data cost. MACs are per window with two flows, including the theory layer for TG-DIN.}
\label{tab:complexity}
\footnotesize
\setlength{\tabcolsep}{4pt}
\begin{tabular}{@{}lrrl@{}}
\toprule
\textbf{Model} & \textbf{Params} & \textbf{MACs} & \textbf{Traffic to collect} \\
\midrule
GRU-LSTM  & 132.3K & 620.8K & 2.7\,h per capacity \\
TFT-style &  75.1K & 354.9K & 2.7\,h per capacity \\
\midrule
TG-DIN    & 102.5K & 171.1K & None (generated) \\
\bottomrule
\end{tabular}
\end{table}

\section{Conclusion}\label{sec:conclusion}

We presented TG-DIN, a theory-guided framework that infers the latent demand flows offer to a shared bottleneck from observable QoS measurements. A differentiable theory layer that encodes bottleneck scheduling and queuing serves two roles: it generates randomized, mechanism-consistent training traces and maps inferred demand to observable throughput, delay, and loss. As a result, TG-DIN learns without demand labels or collected traffic traces, while bottleneck capacity enters explicitly as an input rather than being learned from data. On synthetic traces, TG-DIN remained substantially more robust to capacity shifts than direct predictors trained at a single capacity, including predictors fine-tuned with target-capacity data. Although trained entirely on generated traces, it also transferred to real testbed measurements without fine-tuning, including dynamic capacity changes and application changes. The current theory layer intentionally uses a simple FIFO, tail-drop bottleneck model, yet this minimal mechanism was sufficient to achieve strong cross-capacity generalization and real-traffic transfer. Because the framework requires only a differentiable theory layer, it can naturally incorporate richer network mechanisms, such as alternative scheduling and active queue-management policies, multi-bottleneck paths, and congestion-control-aware traffic generation. Evaluating such extensions on larger and more diverse deployments is a promising direction for future work. More broadly, the inferred demand, expressed in the same rate domain as network capacity, may serve as a useful representation for downstream tasks such as congestion diagnosis, resource allocation, and capacity planning.

\bibliographystyle{IEEEtran}
\bibliography{reference}

@inproceedings{medina2002traffic,
author = {Medina, A. and Taft, N. and Salamatian, K. and Bhattacharyya, S. and Diot, C.},
title = {Traffic matrix estimation: existing techniques and new directions},
year = {2002},
isbn = {158113570X},
publisher = {Association for Computing Machinery},
address = {New York, NY, USA},
doi = {10.1145/633025.633041},
booktitle = {Proceedings of the 2002 Conference on Applications, Technologies, Architectures, and Protocols for Computer Communications},
pages = {161–174},
numpages = {14},
location = {Pittsburgh, Pennsylvania, USA},
series = {SIGCOMM '02}
}

@inproceedings {perry2023dote,
author = {Yarin Perry and Felipe Vieira Frujeri and Chaim Hoch and Srikanth Kandula and Ishai Menache and Michael Schapira and Aviv Tamar},
title = {{DOTE}: Rethinking (Predictive) {WAN} Traffic Engineering},
booktitle = {20th USENIX Symposium on Networked Systems Design and Implementation (NSDI 23)},
year = {2023},
isbn = {978-1-939133-33-5},
address = {Boston, MA},
pages = {1557--1581},
url = {https://www.usenix.org/conference/nsdi23/presentation/perry},
publisher = {USENIX Association},
month = apr
}

@inproceedings{tune2015spatiotemporal,
author = {Tune, Paul and Roughan, Matthew},
title = {Spatiotemporal Traffic Matrix Synthesis},
year = {2015},
isbn = {9781450335423},
publisher = {Association for Computing Machinery},
address = {New York, NY, USA},
doi = {10.1145/2785956.2787471},
booktitle = {Proceedings of the 2015 ACM Conference on Special Interest Group on Data Communication},
pages = {579–592},
numpages = {14},
location = {London, United Kingdom},
series = {SIGCOMM '15}
}

@inproceedings {madhyastha2006iplane,
author = {Harsha Madhyastha and Tomas Isdal and Michael Piatek and Colin Dixon and Thomas Anderson and Arvind Krishnamurthy and Arun Venkataramani},
title = {{iPlane}: An Information Plane for Distributed Services},
booktitle = {7th USENIX Symposium on Operating Systems Design and Implementation (OSDI 06)},
year = {2006},
address = {Seattle, WA},
url = {https://www.usenix.org/conference/osdi-06/iplane-information-plane-distributed-services},
publisher = {USENIX Association},
month = nov
}

@inproceedings{madhyastha2009iplane,
author = {Madhyastha, Harsha V. and Katz-Bassett, Ethan and Anderson, Thomas and Krishnamurthy, Arvind and Venkataramani, Arun},
title = {iPlane Nano: path prediction for peer-to-peer applications},
year = {2009},
publisher = {USENIX Association},
address = {USA},
booktitle = {Proceedings of the 6th USENIX Symposium on Networked Systems Design and Implementation},
pages = {137–152},
numpages = {16},
location = {Boston, Massachusetts},
series = {NSDI'09}
}

@inproceedings{zhang2006measurement,
author = {Zhang, Bo and Ng, T. S. Eugene and Nandi, Animesh and Riedi, Rudolf and Druschel, Peter and Wang, Guohui},
title = {Measurement based analysis, modeling, and synthesis of the internet delay space},
year = {2006},
isbn = {1595935614},
publisher = {Association for Computing Machinery},
address = {New York, NY, USA},
doi = {10.1145/1177080.1177091},
booktitle = {Proceedings of the 6th ACM SIGCOMM Conference on Internet Measurement},
pages = {85–98},
numpages = {14},
location = {Rio de Janeriro, Brazil},
series = {IMC '06}
}

@inproceedings{balachandran2013developing,
author = {Balachandran, Athula and Sekar, Vyas and Akella, Aditya and Seshan, Srinivasan and Stoica, Ion and Zhang, Hui},
title = {Developing a predictive model of quality of experience for internet video},
year = {2013},
isbn = {9781450320566},
publisher = {Association for Computing Machinery},
address = {New York, NY, USA},
doi = {10.1145/2486001.2486025},
booktitle = {Proceedings of the ACM SIGCOMM 2013 Conference on SIGCOMM},
pages = {339–350},
numpages = {12},
location = {Hong Kong, China},
series = {SIGCOMM '13}
}

@inproceedings {iyer2022performance,
author = {Rishabh Iyer and Katerina Argyraki and George Candea},
title = {Performance Interfaces for Network Functions},
booktitle = {19th USENIX Symposium on Networked Systems Design and Implementation (NSDI 22)},
year = {2022},
isbn = {978-1-939133-27-4},
address = {Renton, WA},
pages = {567--584},
url = {https://www.usenix.org/conference/nsdi22/presentation/iyer},
publisher = {USENIX Association},
month = apr
}

@inproceedings {zhao2023scalable,
author = {Kevin Zhao and Prateesh Goyal and Mohammad Alizadeh and Thomas E. Anderson},
title = {Scalable Tail Latency Estimation for Data Center Networks},
booktitle = {20th USENIX Symposium on Networked Systems Design and Implementation (NSDI 23)},
year = {2023},
isbn = {978-1-939133-33-5},
address = {Boston, MA},
pages = {685--702},
url = {https://www.usenix.org/conference/nsdi23/presentation/zhao-kevin},
publisher = {USENIX Association},
month = apr
}

@inproceedings{zhang2023latenseer,
author = {Zhang, Yazhuo and Isaacs, Rebecca and Yue, Yao and Yang, Juncheng and Zhang, Lei and Vigfusson, Ymir},
title = {LatenSeer: Causal Modeling of End-to-End Latency Distributions by Harnessing Distributed Tracing},
year = {2023},
isbn = {9798400703874},
publisher = {Association for Computing Machinery},
address = {New York, NY, USA},
doi = {10.1145/3620678.3624787},
booktitle = {Proceedings of the 2023 ACM Symposium on Cloud Computing},
pages = {502–519},
numpages = {18},
location = {Santa Cruz, CA, USA},
series = {SoCC '23}
}

@inproceedings{carofiglio2021characterizing,
author = {Carofiglio, Giovanna and Grassi, Giulio and Loparco, Enrico and Muscariello, Luca and Papalini, Michele and Samain, Jacques},
title = {Characterizing the Relationship Between Application QoE and Network QoS for Real-Time Services},
year = {2021},
isbn = {9781450386333},
publisher = {Association for Computing Machinery},
address = {New York, NY, USA},
doi = {10.1145/3472727.3472800},
booktitle = {Proceedings of the ACM SIGCOMM 2021 Workshop on Network-Application Integration},
pages = {20–25},
numpages = {6},
location = {Virtual Event, USA},
series = {NAI'21}
}

@inproceedings{jain2013b4,
author = {Jain, Sushant and Kumar, Alok and Mandal, Subhasree and Ong, Joon and Poutievski, Leon and Singh, Arjun and Venkata, Subbaiah and Wanderer, Jim and Zhou, Junlan and Zhu, Min and Zolla, Jon and H{\"o}lzle, Urs and Stuart, Stephen and Vahdat, Amin},
title = {B4: experience with a globally-deployed software defined wan},
year = {2013},
isbn = {9781450320566},
publisher = {Association for Computing Machinery},
address = {New York, NY, USA},
doi = {10.1145/2486001.2486019},
pages = {3–14},
numpages = {12},
location = {Hong Kong, China},
series = {SIGCOMM '13}
}

@inproceedings{hong2013achieving,
author = {Hong, Chi-Yao and Kandula, Srikanth and Mahajan, Ratul and Zhang, Ming and Gill, Vijay and Nanduri, Mohan and Wattenhofer, Roger},
title = {Achieving high utilization with software-driven WAN},
year = {2013},
isbn = {9781450320566},
publisher = {Association for Computing Machinery},
address = {New York, NY, USA},
doi = {10.1145/2486001.2486012},
booktitle = {Proceedings of the ACM SIGCOMM 2013 Conference on SIGCOMM},
pages = {15–26},
numpages = {12},
location = {Hong Kong, China},
series = {SIGCOMM '13}
}

@inproceedings{xu2023teal,
author = {Xu, Zhiying and Yan, Francis Y. and Singh, Rachee and Chiu, Justin T. and Rush, Alexander M. and Yu, Minlan},
title = {Teal: Learning-Accelerated Optimization of WAN Traffic Engineering},
year = {2023},
isbn = {9798400702365},
publisher = {Association for Computing Machinery},
address = {New York, NY, USA},
doi = {10.1145/3603269.3604857},
pages = {378–393},
numpages = {16},
location = {New York, NY, USA},
series = {ACM SIGCOMM '23}
}

@inproceedings{alqiam2024transferable,
author = {AlQiam, Abd AlRhman and Yao, Yuanjun and Wang, Zhaodong and Ahuja, Satyajeet Singh and Zhang, Ying and Rao, Sanjay G. and Ribeiro, Bruno and Tawarmalani, Mohit},
title = {Transferable Neural WAN TE for Changing Topologies},
year = {2024},
isbn = {9798400706141},
publisher = {Association for Computing Machinery},
address = {New York, NY, USA},
doi = {10.1145/3651890.3672237},
booktitle = {Proceedings of the ACM SIGCOMM 2024 Conference},
pages = {86–102},
numpages = {17},
location = {Sydney, NSW, Australia},
series = {ACM SIGCOMM '24}
}

@inproceedings{gui2024redte,
author = {Gui, Fei and Wang, Songtao and Li, Dan and Chen, Li and Gao, Kaihui and Min, Congcong and Wang, Yi},
title = {RedTE: Mitigating Subsecond Traffic Bursts with Real-time and Distributed Traffic Engineering},
year = {2024},
isbn = {9798400706141},
publisher = {Association for Computing Machinery},
address = {New York, NY, USA},
doi = {10.1145/3651890.3672231},
booktitle = {Proceedings of the ACM SIGCOMM 2024 Conference},
pages = {71–85},
numpages = {15},
location = {Sydney, NSW, Australia},
series = {ACM SIGCOMM '24}
}

@inproceedings{zhang2003information,
author = {Zhang, Yin and Roughan, Matthew and Lund, Carsten and Donoho, David},
title = {An information-theoretic approach to traffic matrix estimation},
year = {2003},
isbn = {1581137354},
publisher = {Association for Computing Machinery},
address = {New York, NY, USA},
doi = {10.1145/863955.863990},
pages = {301–312},
numpages = {12},
location = {Karlsruhe, Germany},
series = {SIGCOMM '03}
}

@ARTICLE{xu2021learning,
  author={Xu, Shenghe and Kodialam, Murali and Lakshman, T. V. and Panwar, Shivendra S.},
  journal={IEEE Open Journal of the Communications Society}, 
  title={Learning Based Methods for Traffic Matrix Estimation From Link Measurements}, 
  year={2021},
  volume={2},
  number={},
  pages={488-499},
  doi={10.1109/OJCOMS.2021.3062636}
}

@INPROCEEDINGS{yuan2023traffic,
  author={Yuan, Xinyu and Qiao, Yan and Zhao, Pei and Hu, Rongyao and Zhang, Benchu},
  booktitle={2023 IEEE Symposium on Computers and Communications (ISCC)}, 
  title={Traffic Matrix Estimation based on Denoising Diffusion Probabilistic Model}, 
  year={2023},
  volume={},
  number={},
  pages={316-322},
  doi={10.1109/ISCC58397.2023.10218016}}

@ARTICLE{qiao2024autotomo,
  author={Qiao, Yan and Wu, Kui and Yuan, Xinyu},
  journal={IEEE/ACM Transactions on Networking}, 
  title={AutoTomo: Learning-Based Traffic Estimator Incorporating Network Tomography}, 
  year={2024},
  volume={32},
  number={6},
  pages={4644-4659},
  doi={10.1109/TNET.2024.3424446}}

@inproceedings{lee2020perceive,
author = {Lee, Jinsung and Lee, Sungyong and Lee, Jongyun and Sathyanarayana, Sandesh Dhawaskar and Lim, Hyoyoung and Lee, Jihoon and Zhu, Xiaoqing and Ramakrishnan, Sangeeta and Grunwald, Dirk and Lee, Kyunghan and Ha, Sangtae},
title = {PERCEIVE: deep learning-based cellular uplink prediction using real-time scheduling patterns},
year = {2020},
isbn = {9781450379540},
publisher = {Association for Computing Machinery},
address = {New York, NY, USA},
doi = {10.1145/3386901.3388911},
pages = {377–390},
numpages = {14},
location = {Toronto, Ontario, Canada},
series = {MobiSys '20}
}

@INPROCEEDINGS{shariff2025traffic,
  author={Shariff, Mohammad Arafath Uddin and Karanam, Venkat Sai Suman Lamba and Ramamurthy, Byrav},
  booktitle={2025 IEEE International Conference on Communications Workshops (ICC Workshops)}, 
  title={Traffic Prediction for Research and Education Networks using an Ensemble GRU-LSTM with Varying Lead Times}, 
  year={2025},
  volume={},
  number={},
  pages={1676-1681},
  doi={10.1109/ICCWorkshops67674.2025.11162466}}

@ARTICLE{kougioumtzidis2025mobile,
  author={Kougioumtzidis, Georgios and Poulkov, Vladimir K. and Lazaridis, Pavlos I. and Zaharis, Zaharias D.},
  journal={IEEE Transactions on Artificial Intelligence}, 
  title={Mobile Network Traffic Prediction Using Temporal Fusion Transformer}, 
  year={2025},
  volume={6},
  number={10},
  pages={2685-2699},
  doi={10.1109/TAI.2025.3556627}}

@article{aouedi2025deep,
author = {Aouedi, Ons and Le, Van An and Piamrat, Kandaraj and Ji, Yusheng},
title = {Deep Learning on Network Traffic Prediction: Recent Advances, Analysis, and Future Directions},
year = {2025},
issue_date = {June 2025},
publisher = {Association for Computing Machinery},
address = {New York, NY, USA},
volume = {57},
number = {6},
issn = {0360-0300},
doi = {10.1145/3703447},
journal = {ACM Comput. Surv.},
month = feb,
articleno = {151},
numpages = {37}
}

@ARTICLE{zhou2023spatial,
  author={Zhou, Jie and Ding, Ding and Wu, Ziteng and Xiu, Yuting},
  journal={IEEE Transactions on Network and Service Management}, 
  title={Spatial Context-Aware Time-Series Forecasting for QoS Prediction}, 
  year={2023},
  volume={20},
  number={2},
  pages={918-931},
  doi={10.1109/TNSM.2023.3250512}}

@ARTICLE{hu2025gacl,
  author={Hu, Shengxiang and Zou, Guobing and Zhang, Bofeng and Wu, Shaogang and Lin, Shiyi and Gan, Yanglan and Chen, Yixin},
  journal={IEEE Transactions on Network and Service Management}, 
  title={GACL: Graph Attention Collaborative Learning for Temporal QoS Prediction}, 
  year={2025},
  volume={22},
  number={4},
  pages={3388-3402},
  doi={10.1109/TNSM.2025.3570464}}

@article{liu2023leaf,
author = {Liu, Shinan and Bronzino, Francesco and Schmitt, Paul and Bhagoji, Arjun Nitin and Feamster, Nick and Crespo, Hector Garcia and Coyle, Timothy and Ward, Brian},
title = {LEAF: Navigating Concept Drift in Cellular Networks},
year = {2023},
issue_date = {September 2023},
publisher = {Association for Computing Machinery},
address = {New York, NY, USA},
volume = {1},
number = {CoNEXT2},
doi = {10.1145/3609422},
journal = {Proc. ACM Netw.},
month = sep,
articleno = {7},
numpages = {24}
}

@inproceedings{INSOMNIA,
author = {Andresini, Giuseppina and Pendlebury, Feargus and Pierazzi, Fabio and Loglisci, Corrado and Appice, Annalisa and Cavallaro, Lorenzo},
title = {INSOMNIA: Towards Concept-Drift Robustness in Network Intrusion Detection},
year = {2021},
isbn = {9781450386579},
publisher = {Association for Computing Machinery},
address = {New York, NY, USA},
doi = {10.1145/3474369.3486864},
pages = {111–122},
numpages = {12},
location = {Virtual Event, Republic of Korea},
series = {AISec '21}
}

@inproceedings{InvestigatingLabellessDriftAdaptation,
author = {Kan, Zeliang and Pendlebury, Feargus and Pierazzi, Fabio and Cavallaro, Lorenzo},
title = {Investigating Labelless Drift Adaptation for Malware Detection},
year = {2021},
isbn = {9781450386579},
publisher = {Association for Computing Machinery},
address = {New York, NY, USA},
doi = {10.1145/3474369.3486873},
booktitle = {Proceedings of the 14th ACM Workshop on Artificial Intelligence and Security},
pages = {123–134},
numpages = {12},
location = {Virtual Event, Republic of Korea},
series = {AISec '21}
}

@inproceedings{FastLearning,
author = {Xavier, Bruno Missi and Martinello, Magnos and Trois, Celio and Alenca, Brenno M. and Rios, Ricardo A.},
title = {Fast Learning Enabled by In-Network Drift Detection},
year = {2024},
isbn = {9798400717581},
publisher = {Association for Computing Machinery},
address = {New York, NY, USA},
doi = {10.1145/3663408.3663427},
booktitle = {Proceedings of the 8th Asia-Pacific Workshop on Networking},
pages = {129–134},
numpages = {6},
location = {Sydney, Australia},
series = {APNet '24}
}

@ARTICLE{parekh1993gps,
  author={Parekh, A.K. and Gallager, R.G.},
  journal={IEEE/ACM Transactions on Networking}, 
  title={A generalized processor sharing approach to flow control in integrated services networks: the single-node case}, 
  year={1993},
  volume={1},
  number={3},
  pages={344-357},
  doi={10.1109/90.234856}}

@misc{cisco-cong-mgmt,
  author       = {{Cisco Systems}},
  title        = {Congestion Management Overview --- {IOS QoS}
                  Configuration Guide},
  howpublished = {Cisco Documentation},
  year         = {2012},
  url          = {https://www.cisco.com/c/en/us/td/docs/ios/qos/configuration/guide/12_2sr/qos_12_2sr_book/congstion_mgmt_oview.html}
}

@ARTICLE{ding2025learning,
  author={Ding, Wenlong and Liu, Libin and Chen, Li and Xu, Hong},
  journal={IEEE Transactions on Network Science and Engineering}, 
  title={Learning for Accelerated Traffic Engineering With Differentiable Network Modeling}, 
  year={2026},
  volume={13},
  number={},
  pages={25-36},
  doi={10.1109/TNSE.2025.3579938}}

@misc{che2024differentiable,
      title={Differentiable Discrete Event Simulation for Queuing Network Control}, 
      author={Ethan Che and Jing Dong and Hongseok Namkoong},
      year={2024},
      eprint={2409.03740},
      archivePrefix={arXiv},
      primaryClass={cs.LG},
      url={https://arxiv.org/abs/2409.03740}, 
}

@techreport{rfc7567,
  author      = {F. Baker and G. Fairhurst},
  title       = {{IETF} Recommendations Regarding Active
                 Queue Management},
  institution = {Internet Engineering Task Force},
  type        = {RFC},
  number      = {7567},
  year        = {2015},
  month       = jul,
  url         = {https://www.rfc-editor.org/rfc/rfc7567}
}

@article{carrascosa2022cloud,
title = {Cloud-gaming: Analysis of Google Stadia traffic},
journal = {Computer Communications},
volume = {188},
pages = {99-116},
year = {2022},
issn = {0140-3664},
doi = {https://doi.org/10.1016/j.comcom.2022.03.006},
author = {Marc Carrascosa and Boris Bellalta},
}

\end{document}